\documentclass{article}

\usepackage[hidelinks]{hyperref}
 
\usepackage{amsmath}
\usepackage{amsthm}
\usepackage{amssymb}
\usepackage[ruled,vlined,linesnumbered]{algorithm2e}
\usepackage{graphicx}
\usepackage{fullpage}
\usepackage[hidelinks]{hyperref}
\usepackage{thmtools,thm-restate}

\newcommand{\ttt}[1]{\texttt{#1}}

\newcommand{\RR}{\mathbb R}

\newcommand{\PP}{P}
\newcommand{\QQ}{Q}
\newcommand{\KO}{K}
\newcommand{\nn}{\mathbf n}

\newcommand{\ee}{\varepsilon}

\newcommand{\comm}[1]{}

\newcommand{\stack}{\Sigma}

\newcommand{\se}{\sigma}
\newcommand{\VV}{\mathcal V}
\newcommand{\cutreg}{F}
\newcommand{\cut}{\mathcal C}
\newcommand{\cutdisk}{\mathcal D}
\newcommand{\start}{s}
\newcommand{\terminal}{t}
\newcommand{\arcA}{A}
\newcommand{\arcB}{B}
\newcommand{\tool}{\cutdisk}

\newcommand{\mydef}{:=}

\newcommand{\round}{\mathcal R}

\newcommand{\dis}[1]{}

\DeclareMathOperator{\Int}{Int}

\newtheorem{theorem}{Theorem}[section]

\newtheorem{lemma}[theorem]{Lemma}

\newtheorem{fact}[theorem]{Fact}
\newtheorem{convention}[theorem]{Convention}

\begin{document}

\title{How to Cut Corners and Get Bounded Convex Curvature\footnote{A
preliminary version of this paper was presented at SoCG 2016~\cite{abrahamsen_et_al:LIPIcs:2016:5896}.}}

\author{Mikkel Abrahamsen\footnote{University of Copenhagen and part of BARC, Basic Algorithms Research Copenhagen, supported by the VILLUM Foundation grant 16582. Mikkel Abrahamsen is supported by Starting Grant 1054-00032B from the Independent Research Fund Denmark under the Sapere Aude research career programme.}
\and Mikkel Thorup\footnotemark[2]}

\date{March 5, 2022}

\maketitle

\begin{abstract}
We describe an algorithm for solving an important geometric problem arising in computer-aided manufacturing.
When cutting away a region from a solid piece of material --- such as steel, wood, ceramics, or plastic --- using a rough tool in a milling machine, sharp convex corners of the region cannot be done properly, but have to be left for finer tools that are more expensive to use.
We want to determine a toolpath that maximizes the use of the rough tool.
In order to formulate the problem in mathematical terms, we introduce the notion of bounded convex curvature.
A region of points in the plane $\QQ$ has \emph{bounded convex curvature} if for any point $x\in\partial\QQ$, there is a unit disk $U$ and $\ee>0$ such that $x\in \partial U$ and all points in $U$ within distance $\ee$ from $x$ are in $\QQ$.
This translates to saying that as we traverse the boundary $\partial\QQ$ with the interior of $\QQ$ on the left side, then $\partial\QQ$ turns to the left with curvature at most $1$.
There is no bound on the curvature where $\partial\QQ$ turns to the right.
Given a region of points $\PP$ in the plane, we are now interested in computing the maximum subset $\QQ\subseteq \PP$ of bounded convex curvature.
The difference in the requirement to left- and right-curvature is a natural consequence of different conditions when machining convex and concave areas of $\QQ$.
We devise an algorithm to compute the unique maximum such set $\QQ$, when the boundary of $\PP$ consists of $n$ line segments and circular arcs of arbitrary radii.
In the general case where $\PP$ may have holes, the algorithm runs in time $O(n^2)$ and uses $O(n)$ space.
If $\PP$ is simply-connected, we describe a faster $O(n\log n)$ time algorithm.
\end{abstract}

\section{Introduction}\label{introSec}

\begin{figure}
\centering
\includegraphics[page=18]{collectFigs.pdf}
\caption{Left: A milling machine.
The model is the Rabbit Mill v3.0 from SourceRabbit, who kindly provided permission to use the picture.
\copyright\ SourceRabbit.
Right: A milling tool. Picture by Rocketmagnet, licensed under CC BY-SA 3.0.}
\label{millFig}
\end{figure}

The motivation for our work comes from the generation of toolpaths for milling machines.
Milling is the process of cutting some specified shape in a piece of material --- such as steel, wood, ceramics, or plastic --- using a milling machine; see Figure~\ref{millFig}.
We first describe the clean and general mathematical problem that we solve and afterwards explain how it relates to milling.

Consider a region $\QQ$ of the plane. 
We say that $\QQ$ has \emph{bounded convex curvature} if for any point $x\in\partial\QQ$, there is a unit disk $U$ and $\ee>0$ such that $x\in \partial U$ and $B_\ee(x)\cap U\subset\QQ$, where $B_r(p)\mydef \{q\in\RR^2\colon \|pq\|<r\}$; see Figure~\ref{compositionFig} (left).
This translates to saying that as we traverse $\partial\QQ$ with the interior of $\QQ$ on the left side, then $\partial\QQ$ turns to the left with curvature at most $1$.
There is no bound on the curvature where it turns to the right, and we may even have
sharp concave corners.
Similarly, we say that $\QQ$ has \emph{bounded concave curvature} if the complement $\QQ^c$ has bounded convex curvature.

\begin{figure}
\centering
\includegraphics[page=19]{collectFigs.pdf}
\caption{Left: The sets $\QQ_1$ and $\QQ_2$ have bounded convex curvature, and then $\QQ_1\cup\QQ_2$ does too.
It is demonstrated that the curvature condition is satisfied at the boundary point $x\in\partial\QQ_1$.
Right: A curvilinear region $\PP$.
Dark gray regions are the complement of $\PP$, which consist of four holes and the unbounded exterior of $\PP$.
The white regions are the subset $\QQ$ of bounded convex curvature.
The regions of $\PP\setminus\QQ$ are light gray, i.e., what we need to remove from $\PP$ to get bounded convex curvature.}
\label{compositionFig}
\end{figure}

An appealing composition property of sets of bounded convex curvature is that if we take two such sets $\QQ_1$ and $\QQ_2$, then the union $\QQ_1\cup\QQ_2$ also has bounded convex curvature, as demonstrated in Figure~\ref{compositionFig} (left).
Thus, if $\PP$ is a region of points and $\QQ\subset\PP$ is a maximal (with respect to inclusion) subset of bounded convex curvature, it follows that $\QQ$ is the unique such subset; for if there was another $\QQ'\subseteq \PP$ of bounded convex curvature that was not contained in $\QQ$, then $\QQ\cup\QQ'$ would be a larger subset of bounded convex curvature, contradicting the maximality of $\QQ$.

The input of our problem is a \emph{curvilinear region} $\PP$.
By this we mean that $\PP$ is a connected and compact set of points in the plane bounded by a finite number $n$ of line segments and circular arcs of arbitrary radii.
This representation is quite common in the practical context of computer-aided manufacturing~\cite{Held2005biarc} and it also has the advantage that the output of our algorithm will anyway have this form if the input is a polygonal region (although the output may not be connected).

We present an algorithm that in $O(n^2)$ time finds a subset $\round(\PP) \subseteq \PP$ of bounded convex curvature.
We prove that the result $\round(\PP)$ contains any set $\QQ \subseteq\PP$ of bounded convex curvature, and hence $\round(\PP)$ is the unique maximum such subset.
See Figure~\ref{compositionFig} (right) for an example.
In the special case where $\PP$ is simply-connected, i.e., has no holes, we show how the algorithm can be implemented such that it takes $O(n\log n)$ time.
Even in this case, the result $\round(\PP)$ may be disconnected.

As we will see, the resulting region $\QQ=\round(\PP)$ will be a collection of curvilinear regions bounded by $O(n)$ line segments and circular arcs in total.
The circular arcs in $\partial\QQ\setminus\partial\PP$ (the ``new'' part of the boundary of $\QQ$) all have unit radius and are convex with respect to $\QQ$.
A very useful property of the boundary $\partial\QQ$ is that all concave arcs and concave vertices are also on the boundary of $\PP$.
Indeed, it is easy to verify that if there is a concave arc or vertex on the boundary of $\QQ$ which is not on the boundary of $\PP$, then $\QQ$ is not maximal.

We will now discuss some differences between the various notions of bounded curvature:
\begin{itemize}
\item
The composition property does not hold for sets of bounded concave curvature because if $\QQ_1$ and $\QQ_2$ have bounded concave curvature, then sharp concave corners can appear in $\QQ_1\cup\QQ_2$.
On the other hand, the intersection $\QQ_1\cap\QQ_2=(\QQ_1^c\cup\QQ_2^c)^c$ does have bounded concave curvature and may not have bounded convex curvature.

\item
As we will explain in more detail in Section~\ref{mathFound}, a crucial property is that if $\QQ$ is the (bounded) region enclosed by a simple closed curve $\partial\QQ$ and $\QQ$ has bounded convex curvature, then $\QQ$ contains a unit disk.
A similar property does not hold for sets of bounded concave curvature --- any disk of radius less than $1$ is a counterexample.

\item
As stated above, it follows from the correctness of the algorithm described in this paper that any curvilinear region $\PP$ contains a unique maximal subset $\round(\PP)$ of bounded convex curvature.
It does not hold in general that $\PP$ contains a unique maximal subset of bounded concave curvature (nor of bounded convex \emph{and} concave curvature).
Instead, it holds that there is a unique minimal \emph{superset} of $\PP$ of bounded concave curvature, which we can express as $\round(\PP^c)^c$.
\end{itemize}

\subsection{Applications in milling}
In the following, we describe different contexts in which the problem of computing the maximum subset of bounded convex curvature appears naturally.
The first author came across the problem when he worked as a software engineer for the company Autodesk (a major provider of CAD/CAM software), developing algorithms for computing toolpaths for milling machines.
He implemented a variant of the algorithm presented in this paper which was of great practical use.
The general problem is that we are given a curvilinear region $S$, which we call a \emph{pocket}.
There is a thin layer of material in $S$ close to the boundary
$\partial S$ of $S$. The goal is to remove that layer without removing anything from outside $S$. We are given a rough
tool, and we want to remove as much as possible of the thin layer, leaving as little as
possible for finer tools that are more expensive to use. The output is
a toolpath for the rough tool consisting of one or more curvilinear
regions. The tool is a disk $\tool$ of radius $r$, where $r$ is
bigger than the width of the layer we wish to remove.  The toolpath is
the path which the center of $\tool$ is following and the tool will remove all material in the area swept by $\tool$ as its center follows the toolpath.
The reason that we only have to handle a thin layer close to the boundary of $S$ is because the area farther from the boundary is removed beforehand by tools
that are less precise since they do not get close to the boundary.
Thus we may assume that all material with distance at least $\delta\leq r$ from the boundary has been removed.
Some material closer to the boundary may also have been removed, but this only makes it easier for our tool to move.

Let $\PP$ be the inwards offset of $S$ by $r$, that is, $\PP$ is the subset of $S$ of points with distance at least $r$ to the pocket boundary $\partial S$; see Figure~\ref{beforeAfterRounding}.
Then $\PP$ is the set of all allowed positions of the center of $\tool$ --- if the center goes outside $\PP$, the tool will cut away material from outside the pocket $S$.
If we had complete control over the tool, then we would be able to remove the material at all points with distance at most $r$ from $\PP$ by letting the tool center traverse the boundary $\partial\PP$ (if $\partial\PP$ consists of more than one cycle, we would traverse the cycles one after one, connected by some transitions that we do not care about here), and this would be the maximum amount of material that could possibly be removed from $S$.
However, there are restrictions on what toolpaths we can follow, e.g., we cannot count on following a toolpath with sharp corners in a precise way.

\begin{figure}
\centering
\includegraphics[page=4]{collectFigs.pdf}
\caption{A pocket bounded by $\partial S$. To the left is shown
the boundary $\partial\PP$ of the inwards offset of $S$ by $r$. 
To the right is shown the boundary $\partial\QQ$ of the maximum subset
with bounded
convex curvature of $\PP$.
The dotted arcs in
the corners show the boundary of the material in $S$
that cannot be removed by $\tool$ using the two toolpaths.}
\label{beforeAfterRounding}
\end{figure}

We are now ready to describe the first application where we
want to compute the maximum subset of bounded
convex curvature.

\paragraph{Application 1: Toolpaths of bounded curvature.}
Assume that the tool can only follow the boundary of a curvilinear region with bounded convex \emph{and} concave curvature.
This means that the curvature must be bounded both when the tool is turning left and right.
A double-sided restriction on the curvature like this is usually simply called \emph{bounded curvature} and has been studied in many papers on robotics and routing problems~\cite{agarwal2002curvature,ahn2012reachability,ayala2015length,lazard2002complexity,lee2000approximation}.
In our setup, we furthermore assume that the tool can turn at least as sharply as its own boundary, that is, we have~$r\geq 1$.

Using our algorithm for bounded convex curvature, we are able to identify the toolpath of bounded curvature that will remove the most material, as follows.
First we compute the above set $\PP$ which is the inwards offset of $S$ by $r$.
The boundary $\partial\PP$ can be computed from the Voronoi diagram of $\partial S$ \cite{held1998voronoi}.
As stated before, the toolpath has to stay inside $\PP$.
We now note that every concave part of $\partial\PP$ has curvature at most $1/r\leq 1$, as the concave parts of $\partial\PP$ stem from inwards offsets of concave parts of $\partial S$.
Hence, $\PP$ has bounded concave curvature.
Next we use our algorithm to find the maximum subset $\QQ=\round(\PP)\subseteq \PP$ of bounded convex curvature; see Figure~\ref{beforeAfterRounding}.
The material cut away as $\tool$ follows $\partial\QQ$ is the unique maximum subset that can be cut out of $S$ using a tool with radius $r\geq 1$ and such that $\QQ$ has bounded convex curvature.
However, all concave arcs and concave vertices of $\partial\QQ$ are also in $\partial\PP$, and $\PP$ has bounded concave curvature, so $\QQ$ has bounded convex \emph{and} concave curvature.
It follows that using the toolpath $\partial\QQ$, we remove the maximum subset of the thin layer of material close to the pocket boundary $\partial S$ while respecting the curvature restriction. \\

\begin{figure}
\centering
\includegraphics[page=5]{collectFigs.pdf}
\caption{
The dashed toolpath $\partial \PP$ is the boundary of the inwards offset of
$S$ by $r$. The dotted segments to the corner $w$
show the alternative way of getting around the corner $v$, instead of using the arc $A$.}
\label{heartExFig}
\end{figure}

\begin{figure}
\centering
\includegraphics[page=12]{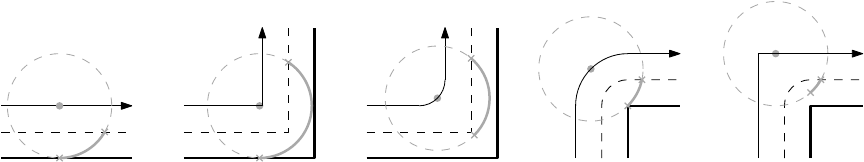}
\caption{In each of these four situations, the thick black curve is the boundary $\partial S$ of the pocket.
The remaining material in the pocket is ensured to be between the dashed black curve and $\partial S$.
The boundary of the tool $\tool$ is the dashed circle, and the solid part of the circle between the two crosses is the maximum part that can be in engagement with the material, i.e., the largest possible portion of the tool boundary cutting away material.
In the third picture, the convex corner on the path in the second picture has been rounded by an arc, thus bounding the convex curvature and reducing the maximum engagement.
The two rightmost pictures show two ways of going around a concave corner of $\partial S$.
In both cases, the maximum engagement is smaller than when the tool
follows a line segment of $\partial S$ (the case of the first picture).}
\label{cuttingCloseups}
\end{figure}

\noindent
In the above example, the set $\PP$ had bounded concave curvature. In
particular, all concave arcs on $\partial\PP$ have radius at least $r$,
and for each concave arc $\arcA$ on $\partial \PP$ of radius
$r$ and center $v$, there is an associated concave vertex $v$ of
$\partial S$. Let $a$ and $b$ be the first and last point on
$\arcA$. When the tool follows $\arcA$, the corner $v$ will be on the
tool boundary $\partial\tool$, and the slightest imprecision will
blunt the corner $v$. A recommended alternative
\cite{park2003mitered} is that we substitute $\arcA$ with two line
segments $aw$ and $wb$ tangential to $\arcA$ at $a$ and $b$,
respectively, thus creating a sharp concave corner $w$ on the
toolpath; see Figures~\ref{heartExFig} and~\ref{cuttingCloseups}. Using this technique, the
corner $v$ will be cut much sharper and more precise. One can think of
various variations of this technique, since we can ``casually'' stop
and turn the tool at any point on its way to $w$ because the remaining
toolpath already ensures that all material will be cut away.  This
shows that we cannot in general assume that there is any bound on the
concave curvature of the input toolpath. We also note the asymmetry with
convex corners and arcs, where overshooting a convex corner implies an
illegal cut through the boundary of $S$.

\paragraph{Application 2: Reducing the load on the tool in convex turns.}
We shall now provide a completely different explanation for the need
for bounded convex curvature. The point is that it is often preferable
for the surface quality of the final product that the tool moves with a constant
speed. Recall that the tool is only removing a thin layer
close to the pocket boundary. The width of this layer is typically a
deliberately chosen fraction of the tool radius $r$.
When moving at constant speed, a convex
turn implies a higher engagement of the tool in the sense of the amount of
material removed per time unit; see Figure~\ref{cuttingCloseups}.
In concave turns the engagement is only decreased.  A too high
engagement could break the tool, and therefore we must bound the
convex curvature of the toolpath.

These and other issues related to the machining of corners have been extensively studied in the technical literature on milling.
See for instance the surveys \cite{hatna1998automatic,toh2004study} and
the papers \cite{burek2019simulation,burek2019numerical,choy2003corner,han2015precise,
iwabe1994study,pateloup2004corner,shixiong2018tool,tong2018prediction,zhao2007pocketing}.
There are several previous papers suggesting methods to get bounded
convex curvature, but none of them guarantees an optimal solution like
ours. One idea for how to handle convex corners is to replace each
of them by a convex circular arc as deep in the corner
as possible. This is suggested and studied
in the papers~\cite{choy2003corner,iwabe1994study,pateloup2004corner,shixiong2018tool}.
However, in all the papers it is assumed that every corner is formed by two
line segments which are
sufficiently long (relative to the angle between them) that a tangential
corner-rounding arc of sufficient size can be placed inside the wedge 
they form.
As can be seen in Figure \ref{beforeAfterRounding}, this is not always the case,
and rounding a toolpath can require more complicated modifications.

\begin{figure}
\centering
\includegraphics[page=8]{collectFigs.pdf}
\caption{The region $\PP$
has bounded convex curvature and all material in $S$ will be machined using $\partial\PP$
as the toolpath.
The innermost dotted curves are the boundary of the inwards offset of $\PP$ by $1$.
The thin corridor of $\PP$ collapses so that the offset is split in two components.
Because of that, the region of $S$ between the thick dashed arcs will not be machined if the double offset method is used.}
\label{boneFig}
\end{figure}

\subsection{Approaches based on Voronoi diagrams}
It may be tempting to think that an algorithm can be obtained by considering the Voronoi diagram of $\PP$.
The Voronoi diagram is a plane graph with one face $f$ for each object $o$ of $\PP$ (where an object is a vertex or an open line segment or circular arc) such that for every point in $f$, the closest object of $\PP$ is $o$.

One heuristic that can be used to obtain a non-trivial subset of bounded convex curvature is the \emph{double offset method}, where we offset $\PP$ inwards by $1$ and then offset the result outwards by $1$ and use that as $\QQ$.
Another way to express the same region is that $\QQ$ is the union of all unit disks contained in $\PP$.
This can be computed from the Voronoi diagram as described by Held, Luk{\'a}cs, and Andor~\cite{held1994pocket}, and using Yap's algorithm~\cite{yap1987ano} for computing Voronoi diagrams, it will take $O(n\log n)$ time.
However, the method does not in general result in the maximum subset of bounded convex curvature.
See for instance Figure~\ref{boneFig}.

\begin{figure}
\centering
\includegraphics[page=17]{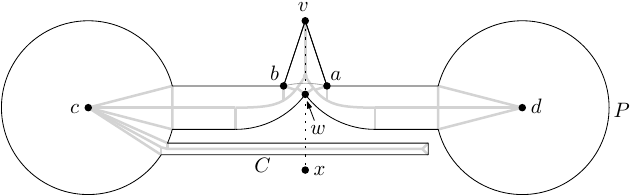}
\caption{An example where it seems that the Voronoi diagram cannot be used to find the maximum subset of bounded convex curvature.
Neither the part of the diagram inside $\PP$ (shown here in fat gray) nor outside $\PP$ contain the center $x$ of the arc from $a$ to $b$.}
\label{fig:considerations2}
\end{figure}

It is in general not true that the Voronoi diagram contains the center of the arcs that are needed to round the parts of the boundary of $\PP$ where the convex curvature is too high.
Consider the example in Figure~\ref{fig:considerations2}.
We assume that the arcs centered at $c$ and $d$ have unit radius.
The Voronoi diagram contains a branch to the convex vertex $v$, and $\PP$ has bounded convex curvature everywhere except at $v$.
To fix this, we need to trim this branch from the Voronoi diagram, but it is not clear how to do this using the Voronoi diagram as the center $x$ of the unit radius arc from $a$ to $b$ is outside $\PP$ and not contained in an edge of the Voronoi diagram (even if extended to the exterior of $\PP$).

Also note the spike with the concave corner $w$.
If the unit radius arc from $a$ to $b$ does not cross the arcs incident at $w$ (as in the figure), then we have obtained the solution by cutting along that arc and removing the corridor $C$.
Otherwise, if $w$ was above the arc from $a$ to $b$, the solution would simply consist of the two unit radius disks with centers $c$ and $d$.
Again, it is not clear how this information can be extracted directly from the Voronoi diagram.

Instead of using the Voronoi diagram, our algorithm works by locally modifying the boundary of $\PP$.
Piece by piece, we remove parts of $\PP$ where the convex curvature is too high.
The difficult part is to do this in such a way that we never remove too much and at the same time terminate after few iterations.

\subsection{Comparison with the conference version}

The paper has been almost completely rewritten since the conference version~\cite{abrahamsen_et_al:LIPIcs:2016:5896}.
First of all, we use another definition of bounded convex curvature, which is shorter, cleaner, and more general than what was used in~\cite{abrahamsen_et_al:LIPIcs:2016:5896}.
In order use this definition, we had to develop a generalized version of the Pestov Ionin Theorem which will be described in Section~\ref{mathFound}.
That work was mathematically challenging and has been published independently~\cite{aam2019disks}, so that this paper can focus on the algorithmic aspects.
Second, we give a more general algorithm that accepts input regions with holes, whereas the input in~\cite{abrahamsen_et_al:LIPIcs:2016:5896} had to be simply-connected.
Third, the nature of the algorithm has changed somewhat:
While we believe the algorithm as described in~\cite{abrahamsen_et_al:LIPIcs:2016:5896} to be correct, we encountered problems when working out subtle technical details in some proofs as we prepared (an earlier version of) this paper.
Our remedy is to describe an algorithm that is slightly less aggressive in the way it cuts away parts of $\PP$ where the convex curvature is too high.
This comes at the cost of a more sophisticated argument bounding the total number of cuts to~$O(n)$.

\subsection{Outline of the paper}
We establish the mathematical foundation of our work in Section~\ref{mathFound} by referring to a recent generalization of the theorem of Pestov and Ionin.
In Section~\ref{algoDescription} we describe an algorithm that computes the maximum subset of bounded convex curvature of $\PP$.
In Section~\ref{anImplementation}, we show that the algorithm has a simple implementation running in time $O(n^2)$ when the input is a general curvilinear region.
We also consider the special case where the input is simply-connected (i.e., a region with no holes), and here we describe a more involved implementation that runs in $O(n\log n)$ time.

\section{Mathematical foundation}\label{mathFound}

\subsection{General notation and conventions}

If $M$ is a set of points in the plane, $\partial M$ denotes the boundary of $M$.
A \emph{curve} $\gamma\colon I\longrightarrow\mathbb R^2$ is a continuous function from some interval $I\subset\RR$ to the plane $\RR^2$.
The curve $\gamma$ is \emph{simple} if it is injective.
We sometimes use $\gamma$ as a short-hand notation for the set of points $\gamma(I)$ on $\gamma$.
A curve $\gamma$ is \emph{closed} if $\gamma\colon [a,b)\to\RR^2$ for some interval $[a,b)$ such that
$\gamma(a)=\lim_{t\nearrow b}\gamma(t)$.

We give simple, closed curves an orientation, either clockwise or counterclockwise.
If $D$ is a disk, we always consider the boundary circle $\partial D$ to have counterclockwise orientation.
If $\gamma$ is a simple, closed curve and $a$ and $b$ are two distinct points on $\gamma$, then $\gamma[a,b]$ denotes the portion of $\gamma$ from $a$ to $b$ in the direction of $\gamma$.
Round parenthesis can be used to exclude one or both endpoints.
If $\gamma$ is not closed, the order of $a$ and $b$ does not matter.
When $\gamma$ is an open curve, we say that a point $c\in\gamma$ is an \emph{inner point} of $\gamma$ if $c$ is not an endpoint of $\gamma$.

\subsection{Curves of bounded convex curvature and the theorem of Pestov and Ionin}

We say that a simple, closed curve $\gamma$ in the plane has \emph{bounded convex curvature} if for every point $x$ on $\gamma$, there is a unit disk $U_x$ and $\ee_x>0$ such that $x\in\partial U_x$ and $B_{\ee_x}(x)\cap U_x\subset\Int\gamma$, where $B_r(p)\mydef \{q\in\RR^2\colon \|pq\|\leq r\}$.
Here, $\Int\gamma$ denotes the region enclosed by $\gamma$.
Note that $\gamma$ has bounded convex curvature if and only if the region $\Int\gamma$ has bounded convex curvature (as defined in Section~\ref{introSec}).
Inspired by the work of the preliminary version of this paper~\cite{abrahamsen_et_al:LIPIcs:2016:5896}, the authors proved the following theorem together with Anders Aamand~\cite{aam2019disks}, generalizing an old result by Pestov and Ionin~\cite{pestov1959largest}.

\begin{theorem}[\cite{aam2019disks}]\label{MAINTHM}
The region enclosed by any curve of bounded convex curvature contains a unit disk.
\end{theorem}

\begin{figure}
\centering
\includegraphics[page=6]{collectFigs.pdf}
\caption{A curve $\gamma$ of bounded convex curvature together with a unit disk $D$ in its interior, the existence of which is guaranteed by Theorem~\ref{MAINTHM}.}
\label{mainThmFig}
\end{figure}

The correctness of the algorithm presented in this paper depends on Theorem~\ref{MAINTHM} as well as the following derived Theorem~\ref{mainThm}.
See Figure~\ref{mainThmFig} for an illustration of the theorem.

\begin{figure}
\centering
\includegraphics[page=7]{collectFigs.pdf}
\caption{Since $V_1\subset R_1$, Theorem~\ref{mainThm} ensures the existence of a unit disk $D\subset R_1$ which is not contained in $V_1$.
The theorem does not make such a promise about $R_2$, since $V_2$ is in the exterior of $R_2$.}
\label{mainThmFig}
\end{figure}

\begin{theorem}\label{mainThm}
Consider a curve $\gamma$ of bounded convex curvature and counterclockwise orientation.
Consider a disk $V$ of any radius, and suppose that for an interval $\gamma[a,b]$ of $\gamma$, it holds that $\gamma[a,b]\cap V=\{a,b\}$.
Let $R$ be the region enclosed by $\gamma[a,b]\cup\partial V(b,a)$.
If $R$ contains $V$, then $R$ contains a unit disk $D$ which is not contained in $V$. 
\end{theorem}

\begin{proof}
The condition that $R$ contains $V$ ensures that $R$ consists of $V$ and the region enclosed by $\gamma[a,b]\cup\partial V(a,b)$, the latter of which is a subset of $\Int\gamma$.
The overall proof is now similar to the one of Theorem~\ref{MAINTHM} from~\cite{aam2019disks}, which we assume the reader to know, but instead of choosing $z$ as an arbitrary point on $\gamma$, we choose $z$ on $\gamma(a,b)$ and choose $D_0$ as the maximal disk tangent to $U_z$ in $z$ and contained in $R$.
If $D_0$ has at least unit radius, we are done.
Otherwise, we observe that $\partial D_0$ meets $\partial R$ in at least two points, none of which can be in $\partial V(b,a)$ since it would imply that $z\in V$.
We can then choose an interval $\gamma(x_0,y_0)\subset\gamma(a,b)$ such that $\gamma[x_0,y_0]\cap\partial D_0=\{x_0,y_0\}$ and proceed as in the original proof.
\end{proof}

The concept of bounded convex curvature is a generalization of previously studied notions of bounded curvature; see~\cite{aam2019disks} for the details.
Versions of the theorem of Pestov and Ionin have often been applied to problems in robot motion planning and related fields~\cite{agarwal2002curvature,ahn2012reachability,ayala2015length,lazard2002complexity,lee2000approximation}.
These papers have studied problems involving curves of bounded curvature, whereas our restriction is one-sided so the class of curves is more general.
Ahn, Cheong, Matou{\v{s}}ek, and Vigneron~\cite[Lemma 2]{ahn2012reachability} gave a result analogous to Theorem~\ref{mainThm} for curves of bounded curvature.

\section{Algorithm}\label{algoDescription}

\subsection{Preliminaries}

We assume that a curvilinear region $\PP_1$ in the plane is given.
That is, $\PP_1$ is a connected, compact region bounded by a finite number of line segments and circular arcs of arbitrary radii.
These form simple, closed curves, where one curve is the outermost and the others bound the holes of $\PP_1$.
The curves bounding the holes are contained in the region enclosed by the outermost curve and the holes have pairwise disjoint interiors.

We set $\PP\mydef \PP_1$ and our algorithm keeps removing parts of $\PP$
while maintaining the invariant that
$\PP$ contains every subset of
$\PP_1$ of bounded convex curvature.
In the end, $\PP$ itself has bounded convex curvature and it follows
that $\PP$ is the unique maximum subset of $\PP_1$
of bounded convex curvature.

The region $\PP$ is always a collection of curvilinear regions.
Each region $\KO$ is represented by the cycles forming the boundary $\partial\KO$.
Each cycle on the boundary $\partial\KO$ is represented as a set of points known as the \emph{vertices}
of $\KO$ (and of $\PP$) and a set of line segments and circular arcs known as the \emph{arcs}
of $\KO$ (and of $\PP$).
An \emph{object} of $\KO$ (and of $\PP$) is a vertex or an arc of $\KO$.

We think of line segments as circular arcs with infinite radius and therefore in most cases we use the word \emph{arcs} for both circular arcs and line segments.
Depending on the context, we may consider a vertex as a point or a set
containing a single point.
We use the convention that an arc includes its endpoints.
Every two arcs of $\PP$ are disjoint except possibly at the endpoints, and
for each vertex there are exactly two arcs incident at that vertex.
This way, the arcs form the closed curves bounding $\PP$.
We always use $n$ to denote the number of vertices of the input $\PP_1$.

Each cycle on the boundary $\partial\PP$ is oriented so that the interior of $\PP$ is on the left.
For an outer cycle, this will be the counterclockwise direction, while for cycles bounding the holes, it will be clockwise.
Similarly, we orient each arc following the orientation of the cycle containing it.
We denote the endpoints of an arc $\arcA$ as $\start(\arcA)$ and $\terminal(\arcA)$, so that $\arcA$ starts at $\start(\arcA)$ and ends at $\terminal(\arcA)$.
An arc $\arcA$ of $\PP$ is \emph{convex} if $\arcA$ turns to the left when traversed in forward direction.
Otherwise, $\arcA$ is \emph{concave}.
If $\arcA$ is a line segment, we regard it as both convex and concave at the same time.
A vertex $v$ is \emph{convex} if the interior angle of $\partial\PP$ at $v$ is strictly less
than $\pi$. If the angle is strictly more than $\pi$, $v$ is \emph{concave}.
If the angle is exactly $\pi$, then $v$ is \emph{tangential}.

When $a$ and $b$ are points on the same boundary cycle $\gamma$ of $\PP$, we define $\partial\PP[a,b]$ and $\partial\KO[a,b]$ to mean the portion of $\gamma$ from $a$ to $b$ with respect to the orientation of $\gamma$.
Again, we may use round parenthesis to exclude one or both endpoints.

We require that no arc of the input $\PP_1$ spans an angle of more than $\pi$.
Arcs that span a larger angle can be split into two arcs connected by a tangential vertex.
This at most doubles the number of vertices and arcs.

Let $\arcA$ be an arc of $\PP$ and $a\in \arcA$ a point on $\arcA$. Then
$\nn_\arcA(a)$ is the unit normal of $\arcA$ at $a$ which points to the left
relative to the orientation of $\arcA$.
We say that two arcs $\arcA$ and $\arcB$ are \emph{tangential} if
$\terminal(\arcA)=\start(\arcB)$ and
$\nn_{\arcA}(\terminal(\arcA))=\nn_{\arcB}(\start(\arcB))$.
Note that $\arcA$ and $\arcB$ are tangential if and only if
the vertex $\terminal(\arcA)$ is tangential.

\subsection{Basic algorithm}

Until Section~\ref{sec:infinite}, we will consider a basic version of our algorithm, which is written as pseudocode in Algorithm~\ref{roundAlgSimple}.
As we will see, the algorithm is correct in the sense that if it returns $\PP$, then $\PP$ is the maximum subset of the input $\PP_1$ of bounded convex curvature.
Unfortunately, the algorithm may go into an infinite loop and never return anything.
In Section~\ref{sec:active}, we will describe a more aggressive algorithm which terminates after $O(n)$ iterations.
The aggressive algorithm is similar to the basic algorithm, but makes some shortcuts to avoid the infinite loops.
We therefore describe the basic algorithm first, as it nicely illustrates the principle behind both algorithms.

The basic format is to repeatedly choose an object $\se$ of $\PP$ that violates the condition of bounded convex curvature.
We say that such an object $\se$ is \emph{problematic}, and it can be a convex arc of radius less than $1$ or a convex vertex. 
In each iteration of the loop at line~\ref{simple:loop}, we pick an arbitrary problematic object $\se$ and eliminate it by removing from $\PP$ a subset $\VV\subseteq \PP$.
By \emph{performing} a \emph{cut}, or simply a \emph{cut}, we mean the process of removing $\VV$ from $\PP$.
It is important to choose $\VV$ such that $\VV\cap\QQ=\emptyset$ for every set $\QQ\subset\PP_1$ of bounded convex curvature.
Theorems~\ref{MAINTHM} and \ref{mainThm} will be used to prove this.

It is possible that a cut splits $\PP$ into more components.
The algorithm will then keep working on each component separately.
Most cuts will introduce one or more new unit radius cut arcs on $\partial \PP$, the endpoints of which can be convex vertices.
The algorithm will proceed to work on these in subsequent iterations.

\begin{algorithm}[h]
\LinesNumbered
\DontPrintSemicolon
\SetArgSty{}
\SetKwInput{Input}{Input}
\SetKwInput{Output}{Output}
\SetKw{Report}{report}
\SetKwIF{If}{ElseIf}{Else}{if}{}{else if}{else}{end if}
\SetKwFor{Foreach}{for each}{}{end for}
\SetKwFor{While}{while}{}{end while}
\While {$\PP$ does not have bounded convex curvature} {\label{simple:loop}
  Let $\se$ be a convex arc of radius less than $1$ or a convex vertex of $\PP$. \\
  Remove $\VV$ from $\PP$. \\
}
\Return {$\PP$}
\caption{$\ttt{BasicBoundedConvexCurvature}(\PP)$}
\label{roundAlgSimple}
\end{algorithm}

Let $\PP_i$ be the set $\PP$ in the beginning of iteration $i\in\{1,2,\ldots\}$ of the loop at line \ref{loop} in Algorithm~\ref{roundAlgSimple}.
Hence we have
\[
\PP_1\supseteq\PP_2\supseteq\cdots.
\]
The challenge is to define $\VV$ in each iteration so that:
\begin{enumerate}
\item
$\PP\mydef\PP\setminus\VV$ can be computed efficiently,
\item
$\VV\cap\QQ=\emptyset$ for every set $\QQ\subset\PP_1$ of bounded convex curvature, and
\item
the total number of iterations is linear, i.e., the algorithm returns $\PP_k$ for some $k=O(n)$.
\end{enumerate}
The last requirement will not be satisfied by the basic algorithm described here.
In Section~\ref{sec:active} we will see that by making the algorithm more aggressive so that it removes larger sets $\VV$ in some cases, we obtain an algorithm that fulfills all requirements.

\subsection{Specifying a cut using an arc $\cut$}\label{sec:specifying}

\begin{figure}
\centering
\includegraphics[page=2]{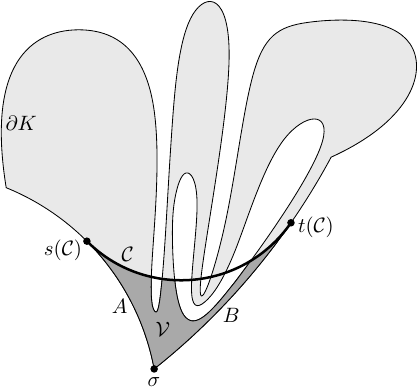}
\caption{
The dark gray area is the region $\VV$ that will be removed in order to handle the convex corner $\se$.
This cut will split $\KO$ into two smaller components, one with a single new cut arc and the other with two.
}
\label{cutFig}
\end{figure}

Let $\se$ be a problematic object of $\PP$ 
that is to be eliminated in a certain iteration of Algorithm~\ref{roundAlgSimple}, and let $\KO$ be the connected component of $\PP$ such that $\se$ is an object of $\KO$.
If we conclude that $\KO$ does not contain any non-empty set of
bounded convex curvature, we set $\VV\mydef \KO$, so that all of the component $\KO$ is removed from
$\PP$.
We may conclude so when $\KO$ has only $O(1)$ arcs.
Otherwise, we specify $\VV=\VV(\cut)$ using a circular arc $\cut$
of unit radius; see Figure \ref{cutFig}.
Assume for now that we have defined $\cut$.
Let $\start(\cut)$ and $\terminal(\cut)$ be the points where $\cut$ starts
and ends in counterclockwise direction, respectively.
The endpoints are included in $\cut$ and
are points on $\partial \KO$.
They will be defined so that $\se\subseteq \partial\KO[\start(\cut),\terminal(\cut)]$.

If $\se$ is a convex vertex, then $\cut$ will connect the arc preceding $\se$ with the arc succeeding $\se$ on $\partial\KO$.
Otherwise, $\se$ is a convex arc of radius less than $1$, and $\cut$ will connect the endpoints of $\se$.
It is furthermore the case that $\cut$ does not intersect $\partial\KO(\start(\cut),\terminal(\cut))$, so that $\cut\cup \partial\KO(\start(\cut),\terminal(\cut))$ forms a simple, closed curve consisting of up to three arcs.
Let $\cutreg(\cut)$ be the region enclosed by this curve.
The arc $\cut$ will be chosen such that it is a concave arc of $\cutreg(\cut)$.

The arc $\cut$ divides $\KO$ into regions $R_1,\ldots,R_r$, which are
the connected components of $\KO\setminus\cut$.
Since we choose $\cut$ such that $\cut\cap \partial\KO(\start(\cut),\terminal(\cut))=\emptyset$, it follows that exactly one of the regions contains $\partial\KO[\start(\cut),\terminal(\cut)]$ on the boundary, and we may assume that it is $R_1$.
We now simply define $\VV\mydef R_1$ as the region to be removed from $\KO$.
Note that $\VV\subseteq\cutreg(\cut)$, but as shown in Figure~\ref{cutFig}, there may also be other regions $R_i$ where $R_i\subset\cutreg(\cut)$, and these we do not remove from $\KO$.

We now have that $\KO \setminus\VV$ is closed and so is our new
$\PP\mydef \PP \setminus \VV$.
The arc $\cut$ will be carefully chosen so that
$\VV\cap\QQ=\emptyset$ for every set $\QQ\subset\PP_1$
of bounded convex curvature.

Consider the situation where we have chosen an arc $\cut$ and thus removed a proper subset $\VV$ from a connected component $\KO$ of $\PP_i$ to obtain $\PP_{i+1}$.
The remaining part of $\KO$ consists of a number $t$ of components $\KO_1,\ldots,\KO_t$, and each of these $\KO_j$ will have one or more arcs $\cut_1,\ldots,\cut_u$ which are contained in $\cut$.
In other words, these arcs are the common boundary of the remaining component $\KO_j$ and the removed part $\VV$.
If $\KO$ is simply-connected, there will be exactly one such arc for each $\KO_j$, but when $\KO$ has holes, there can be more, as seen in Figure~\ref{cutFig}.
We denote the arcs $\cut_{1},\ldots,\cut_{u}$ as \emph{new cut arcs} of $\PP_{i+1}$.
Note that since the arc $\cut$ defining a cut is a concave arc of $\cutreg(\cut)$, $\VV\subseteq\cutreg(\cut)$, and the common boundary between $\PP_{i+1}$ and $\VV$ consists of arcs on $\cut$, we get that all new cut arcs of $\PP_{i+1}$ are convex.

In general, a \emph{cut arc} of $\PP_{i+1}$ is a convex arc of unit radius. 
For technical reasons, this even includes convex arcs of unit radius of the original input $\PP_1$.
We say that a cut arc $\arcA$ is \emph{perfect} if no endpoint of $\arcA$ is a convex vertex.

An important concept in our algorithm is that of a unit radius arc from one point to another.
Consider two points $p$ and $q$ with distance at most $2$ and let $D$ be the unique unit disk such that $p,q\in \partial D$ and the arc $\partial D[p,q]$ spans an angle of at most $\pi$.
We then define the \emph{unit radius arc} from $p$ to $q$ to be this arc $\partial D[p,q]$.
In fact, it will always hold that $\cut$ is the unit radius arc from $\start(\cut)$ to $\terminal(\cut)$.

Recall that we require that all arcs of the input $\PP_1$ span an angle of at most $\pi$.
We now show that this property is maintained by the algorithm.

\begin{lemma}\label{lem:leqpi}
Every arc created by the algorithm spans an angle of at most $\pi$.
\end{lemma}

\begin{proof}
The only arcs for which we need to verify the property are the new cut arcs, as all other arcs can only shrink.
The statement thus holds if for every cut we make, the arc $\cut$ specifying the cut spans an angle of less than $\pi$.
This is trivial since we define $\cut$ as the unit radius arc from a point $p$ to a point $q$, which by definition is an arc spanning an angle of at most $\pi$.
\end{proof}

Before we can describe how to choose the arcs defining our cuts in Section~\ref{cutDef}, we need the notions of touching disks and bisectors of arcs, as defined in the following section.

\subsection{Touching disks and bisectors of arcs}

Let $\arcA$ be an arc of $\PP$.
A disk $C$ \emph{touches} $\arcA$ at $a\in \arcA$ if
\begin{itemize}
\item
the interior of $C$ is disjoint from $\arcA$,

\item
$\partial C$ and $\arcA$ have the
point $a$ in common, and

\item
the center $x$ of $C$ is the point $a+c\cdot \nn_\arcA(a)$ where $c\geq 0$ is the radius of $C$ (recall that $\nn_\arcA(a)$ is the unit normal to $A$ in $a$ pointing to the left).
\end{itemize}

\begin{figure}
\centering
\includegraphics[page=13]{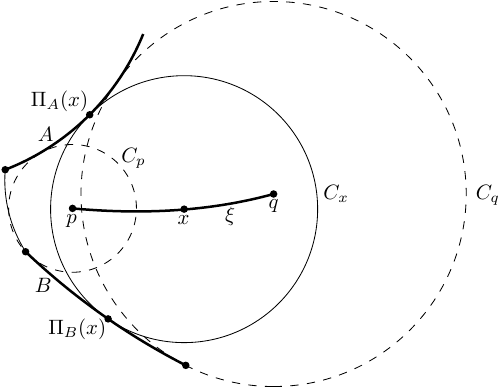}
\caption{
Two arcs $\arcA$ and $\arcB$ and the bisector $\xi=\xi(\arcA,\arcB)$ that they define.
Here, $\xi$ is limited by $\arcB$.
}
\label{fig:bisector}
\end{figure}

We allow the case $c=0$, where $C$ is a single point.
Note that for $C$ to touch an arc $\arcA$ of $\PP$, we do not require that $C$ be contained in $\PP$.

It always holds that the point $a$ where $C$ touches $\arcA$ is the point on $\arcA$ closest to the center $x$ of $C$.
A special case happens when $\arcA$ is an arc on $\partial C$, in which case $\arcA$ is a convex arc and $C$ touches $\arcA$ at every point on $\arcA$.

The \emph{bisector} $\xi\mydef \xi(\arcA,\arcB)$ of two arcs $\arcA$ and $\arcB$ of $\PP$ is the locus of the centers of all disks that touch both $\arcA$ and $\arcB$; see Figure~\ref{fig:bisector}.
Then $\xi$ is an open curve, and Yap~\cite{yap1987ano} showed that $\xi$ is part of a conic section.
For a point $x\in\xi$, we define $C_x=C_x(\arcA,\arcB)$ to be the disk with center $x$ that touches $\arcA$ and $\arcB$.
Let the endpoints of $\xi$ be $p$ and $q$ such that $C_p$ touches $\arcA$ at $\terminal(\arcA)$ or $\arcB$ at $\start(\arcB)$, and $C_q$ touches $\arcA$ at $\start(\arcA)$ or $\arcB$ at $\terminal(\arcB)$.
We then orient $\xi$ from $p$ to $q$.
If $C_q$ contains $\start(\arcA)$, we say that $\arcA$ \emph{limits} $\xi$, and otherwise that $\arcB$ \emph{limits} $\xi$.
Note that sometimes $\xi$ is not be contained in $\PP$.

For a point $x\in\xi$, we define $\Pi_{\arcA}(x)=\Pi_{\arcA}(\arcA,\arcB,x)$ as the point $C_x\cap \arcA$.
In the exceptional case where $x$ is the center of $\arcA$ so that this is not unique, we define $\Pi_{\arcA}(x)=\start(\arcA)$.
We define $\Pi_{\arcB}(x)$ similarly, except that when $x$ is the center of $\arcB$, we define $\Pi_{\arcB}(x)=\terminal(\arcB)$.

When defining the arc $\cut$ that specifies the cut we want to perform, we are often searching for the first point $x$ on a bisector $\xi$, with respect to the orientation of $\xi$, where $C_{x}$ has radius $1$.
If there is no such point, we choose $x$ as the end of $\xi$.
Using elementary geometry, we can compute the point $x$ in constant time.

When we have chosen $x$, we shall define $\cut$ to be the unit radius arc from $\Pi_{\arcA}(x)$ to $\Pi_{\arcB}(x)$.
If $C_x$ has radius $1$, then $\cut$ will be an arc on $\partial C_x$, but otherwise, the radius of $C_x$ is less than $1$ and then $\cut$ will be passing through the interior of $C_x$.

Since $C_x$ touches $\arcA$ and $\arcB$, we have that $C_x\cap \arcA$ and $C_x\cap\arcB$ consist of the single points $\Pi_\arcA(x)$ and $\Pi_\arcB(x)$, except in the special case where $\arcA$ or $\arcB$ is an arc on $\partial C_x$.
Since $\cut$ is contained in $C_x$, we can therefore likewise conclude that $\cut\cap\arcA=\Pi_\arcA(x)$ and $\cut\cap\arcB=\Pi_\arcB(x)$, unless $\arcA$ and $\cut$ or $\arcB$ and $\cut$ are both arcs on $\partial C_x$.
This property will be crucial when proving that the algorithm works, since it allows us to prove that $\cut\cup \partial\PP(\start(\cut),\terminal(\cut))$ forms a simple, closed curve.

\subsection{Defining the arc $\cut$ specifying a cut}\label{cutDef}

The arc $\cut$ that defines a cut is specified by one of the following two types.
Recall that $\se$ is an object of $\PP$, which is either a convex arc of radius less than $1$ or a convex vertex. 
Figure~\ref{fig:cuttypesSimple} shows an example of each type of cut.
In each case, we define $\arcA$ to be the arc preceding $\se$ and $\arcB$ to be the arc succeeding $\se$ on $\partial\PP$.
The statement inside the square brackets [ ] specifies when the type of cut applies.

\begin{figure}
\centering
\includegraphics[page=11]{collectFigs.pdf}
\caption{An example of each type of cut of the basic algorithm.}
\label{fig:cuttypesSimple}
\end{figure}

\begin{enumerate}
\item[1.]
[$\se$ is a convex arc with radius less than $1$.]
We define $\cut$ as the unit radius arc from $\start(\se)$ to $\terminal(\se)$.

\item[2.]
[$\se$ is a convex vertex of $\PP$.]
We consider the bisector $\xi\mydef \xi(\arcA,\arcB)$.
 
\begin{enumerate}
\item[2.1.]
[There is $x\in\xi$ such that $C_x$ has radius $1$.]
We choose the first such $x$ and define $\cut$ as the unit radius arc from $\Pi_{\arcA}(x)$ to $\Pi_{\arcB}(x)$, which is the same as the arc $C_x[\Pi_{\arcA}(x),\Pi_{\arcB}(x)]$.

\item[2.2.]
[Otherwise], we choose $x$ as the end of $\xi$ and $\cut$ as the unit radius arc from $\Pi_{\arcA}(x)$ to $\Pi_{\arcB}(x)$.
\end{enumerate}
\end{enumerate}

This finishes the description of the basic algorithm.

\subsection{Basic properties of cuts}\label{sec:basicSimple}

In this section, we prove that the way we defined the arcs specifying cuts in Section~\ref{cutDef} ensures that the cuts have certain important  properties.
Recall that we claimed the set $\cut\cup \partial\PP(\start(\cut),\terminal(\cut))$ to be a simple, closed curve, and we defined $\cutreg(\cut)$ to be the region bounded by this curve.
This property was used when defining the set $\VV$ to be removed in each cut.
We now prove this claim.

\begin{restatable}{lemma}{simplecurvelemma}
\label{lem:simplecurve}
The arc $\cut$ intersects $\partial\PP[\start(\cut),\terminal(\cut)]$ at the points $\start(\cut)$ and $\terminal(\cut)$ only.
Thus, it holds that $\cut\cup \partial\PP(\start(\cut),\terminal(\cut))$ is a simple, closed curve.
\end{restatable}

\begin{proof}
For cuts of type 1, the arcs $\cut$ and $\sigma=\partial\PP[\start(\cut),\terminal(\cut)]$ intersect at $\start(\cut)$ and $\terminal(\cut)$.
There can be at most $2$ intersection points as the arcs have different radii, so the statement holds in this case.

Now consider a cut of type 2.
Here, we choose $\cut$ in the disk $C_x$ that touches $\arcA$ and $\arcB$.
We claim that $C_x\cap \arcA$ consists of a single point, and similarly for $C_x\cap \arcB$.
If $C_x\cap \arcA$ consists of more than one point, then it must be the case that $\arcA$ is an arc on $C_x$.
Now, since $\arcA$ and $\arcB$ have the point $\se=\terminal(\arcA)=\start(\arcB)$ in common, $C_x$ also touches $\arcB$ at $\se$.
Hence, $\arcA$ and $\arcB$ are tangential at $\se$, which contradicts this type of cut.
As $C_x\cap \arcA$ and $C_x\cap \arcB$ consist of single points and $\cut$ is contained in $C_x$, the statement now follows.
\end{proof}

We define $\cutdisk=\cutdisk(\cut)$ to be the unit disk with $\cut$ on the boundary.
As we will see, the correctness of the algorithm relies on the following crucial lemma.

\begin{restatable}{lemma}{unitdisklemma}
\label{lem:unitpropSimple}
The only unit disk contained in $\cutreg(\cut)\cup \cutdisk$ is $\cutdisk$.
\end{restatable}

\begin{proof}
If there is another unit disk $E$ in $\cutreg(\cut)\cup \cutdisk$, then $E$ overlaps $F(\cut)$.
For a cut of type 1, that is clearly not possible as $\se$ is a convex arc of radius less than $1$.

Let us now consider a cut of type 2.1; a cut of type 2.2 can be analyzed in a similar way.
We move $E$ in the direction away from the center of $\cutdisk$ until we are blocked by the boundary of $\cutreg(\cut)\cup \cutdisk$, so that there is a point $p\in\partial E\cap\partial (\cutreg(\cut)\cup \cutdisk)$.
The proof now branches out in several cases depending on where this point $p$ is on $\partial (\cutreg(\cut)\cup \cutdisk)$, and all of these cases are easy to refute.
 
Suppose first that $p\in \cutdisk(\terminal(\cut),\start(\cut))$.
Since $\cutdisk$ and $E$ are both disks of radius $1$, this implies that they are the same disk, which is a contradiction.

Suppose now that $p\in\{\start(\cut),\terminal(\cut)\}$.
Since $\cutdisk$ touches $\arcA$ and $\arcB$ in these points and $\cutdisk$ and $E$ both have radius $1$, it again follows that they must be the same disk, which is a contradiction.

Therefore, we are left with the possibility that $p\in \partial \PP(\start(\cut),\terminal(\cut))$.
Hence, we have $p\in \arcA(\start(\cut),\se]$ or $p\in\arcB[\se,\terminal(\cut))$.
It follows that there exists a unit disk osculating $\arcA$ and $\arcB$ at points closer to $\se$ than $\cutdisk$ does, which is a contradiction.
\end{proof}

With these lemmas, we are now ready to prove that the basic algorithm is correct.

\subsection{Correctness of the basic algorithm}\label{correctness}

\begin{figure}
\centering
\includegraphics[page=22]{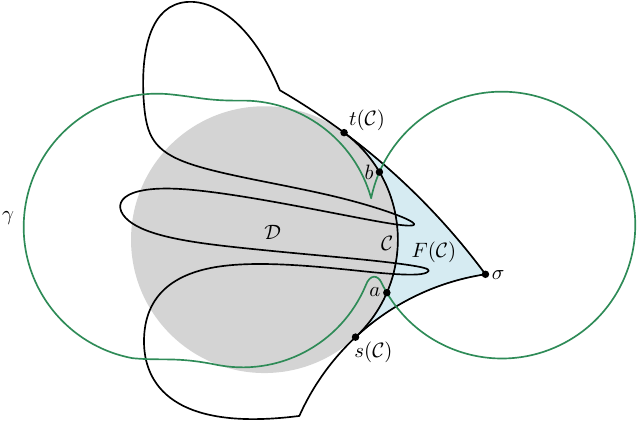}
\caption{Lemma~\ref{lemma:algCurve} says that the region $\cutreg(\cut)$ does not contain a curve of bounded convex curvature, such as the curve $\gamma$ shown here.
Furthermore, the interval $\gamma[a,b]$ cannot be contained in $\cutreg(\cut)$, as $\cutdisk$ is contained in the region bounded by $\gamma[a,b]\cup\partial\cutdisk(b,a)$.}
\label{fig:lemma:algCurve}
\end{figure}

In this section, we prove that the set $\VV$ to be cut away from $\PP$
is disjoint from every subset of $\PP_1$ of bounded convex curvature.
The proof is by contradiction. The idea is that if there is a subset of
$\PP_1$ of bounded convex curvature
that overlaps $\VV$, then Theorem~\ref{MAINTHM} or~\ref{mainThm} shows the existence of a unit disk in $\VV$ which cannot be there according to Lemma~\ref{lem:unitpropSimple}.
We first prove a lemma of a more general nature, which will also turn out useful when describing an implementation of our algorithm in Section~\ref{anImplementation}; see Figure~\ref{fig:lemma:algCurve}.

\begin{restatable}{lemma}{lemmaalgcurve}
\label{lemma:algCurve}
Let $\cut$ be the arc that defines the cut we make to get from $\PP_i$ to $\PP_{i+1}$ for any iteration $i$.
For every curve $\gamma$ of bounded convex curvature and counterclockwise orientation, the following holds.
\begin{enumerate}
\item
$\gamma\not\subset \cutreg(\cut)$.

\item\label{lemma:algCurve:it2}
Suppose that an interval $\gamma[a,b]\subset\gamma$ satisfies $a,b\in\cut$, $\gamma[a,b]\cap \PP_{i+1}=\{a,b\}$, and that the region bounded by $\gamma[a,b]\cup\partial\cutdisk(b,a)$ contains $\cutdisk=\cutdisk(\cut)$.
Then $\gamma[a,b]\not\subset \cutreg(\cut)$.
\end{enumerate}
\end{restatable}

\begin{proof}
Consider for contradiction
a situation where the statement is violated by a curve $\gamma$ of bounded convex curvature.
Suppose first that $\gamma\subset \cutreg(\cut)$.
By Theorem~\ref{MAINTHM}, the interior of $\gamma$ contains a unit disk $E$ which is also contained in $\cutreg(\cut)$, but that contradicts Lemma~\ref{lem:unitpropSimple}.

Consider now an interval $\gamma[a,b]$ as described in point~\ref{lemma:algCurve:it2}.
Assume for contradiction that $\gamma[a,b]\subset \cutreg(\cut)$.
By Theorem~\ref{mainThm}, we get that the region $R$ bounded by $\gamma[a,b]$ and $\partial\cutdisk(b,a)$ contains a unit disk different from $\cutdisk$.
As $R\subset\cutreg(\cut)\cup\cutdisk$, this contradicts Lemma~\ref{lem:unitpropSimple}.
\end{proof}

We are now ready to prove the correctness of the simple algorithm.

\begin{restatable}{theorem}{thmcorrect}
\label{algOptimalSimple}
For every set $\QQ\subset \PP_1$ of bounded convex curvature and every iteration $i$, the set $\PP_i$ contains~$\QQ$.
\end{restatable}

\begin{proof}
Let $\QQ\subset \PP_1$ be a set of bounded convex curvature, and assume for contradiction that $\QQ\not\subset\PP_{i+1}$ while $\QQ\subset\PP_i$, where $i$ is chosen to be minimal.
Let $\cut$ be the arc defining the cut we make in $\PP_i$ in order to get $\PP_{i+1}$.
Without loss of generality, we can assume that $\QQ$ has one connected component.
The outer boundary of $\QQ$ is a curve $\gamma$ of bounded convex curvature.
If $\gamma\subset(\PP_1\setminus\PP_{i+1})$, we get from the minimality of $i$ that also $\gamma\subset(\PP_i\setminus\PP_{i+1})=\VV(\cut)$.
Since $\VV(\cut)\subset \cutreg(\cut)$ by definition, this contradicts Lemma~\ref{lemma:algCurve}.

On the other hand, if $\gamma\not\subset(\PP_1\setminus\PP_{i+1})$, we get by the minimality of $i$ that $\gamma$ exits $\PP_{i+1}$ through a point on $\cut$.
Let $a$ be the point in $\cut\cap \gamma$ closest to $\start(\cut)$.
We now follow $\gamma$ after $a$ and observe that $\gamma$ must enter $\PP_{i+1}$ at a point $b$ on $\cut(a,\terminal(\cut)]$.
Hence, the region $R$ bounded by $\gamma[a,b]$ and $\partial\cutdisk(b,a)$ contains $\cutdisk=\cutdisk(\cut)$.
By the minimality of $i$, we have $\gamma[a,b]\subset \VV(\cut)\subset\cutreg(\cut)$.
However, the existence of such an interval $\gamma[a,b]$ contradicts Lemma~\ref{lemma:algCurve}.
\end{proof}

\subsection{Infinite loops}\label{sec:infinite}

\begin{figure}
\centering
\includegraphics[page=16]{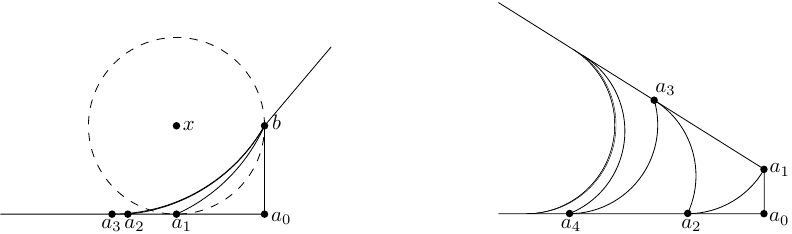}
\caption{Left: The basic algorithm can end in an infinite loop of cuts of type 2.2 when rounding vertex $a_0$.
The algorithm would create convex vertices $a_1,a_2,\ldots$ that converge to a tangential vertex.
Right: Even if we make a more sophisticated type of cut where the arc $\cut$ is tangential to one arc and goes to the endpoint of the other, we can get into an infinite loop.
In order to round $a_0$, we would create and thereafter round the infinite sequence of vertices $a_1,a_2,\ldots$.}
\label{fig:considerations1}
\end{figure}

Figure~\ref{fig:considerations1} (left) shows that the basic algorithm may go into an infinite loop.
Here, the algorithm keeps making cuts of type 2.2, because such a cut always produces a new convex vertex (although in the limit, a tangential vertex is created).
If we are slightly more careful, we can instead of type 2.2 introduce new types of cuts:
When type 2.1 is not applicable, we first try to use an arc which is tangential to $\arcA$ or $\arcB$ and has an endpoint in $\terminal(\arcB)$ or $\start(\arcA)$, respectively.
If that is not possible, we choose $\cut$ as the unit radius arc from $\start(\arcA)$ to $\terminal(\arcB)$.
With these rules, we would avoid the infinite loop seen in Figure~\ref{fig:considerations1} (left), as we would jump directly to the arc in the limit.
However, as shown in Figure~\ref{fig:considerations1} (right), it would still be possible for the algorithm to go into an infinite loop.
Due to such phenomena, we introduce the concept of an \emph{active} cut arc, which is a cut arc introduced in an earlier iteration where one or both endpoint are convex vertices of $\PP$, as explained in the next section.

\subsection{Active cut arcs and aggressive algorithm}\label{sec:active}

As we saw in the previous section, the basic algorithm may make an infinite number of iterations where cut arcs are created with a convex endpoint, such that we never obtain a set of bounded convex curvature.
In order to resolve this issue, we introduce the notion of an \emph{active} cut arc, which is a new type of problematic object that our algorithm must be able to handle (the other two types being convex vertices and convex arcs of radius less than $1$, as in the basic algorithm).

The active cut arcs get priority to be handled before the other problematic objects in the same component of $\PP$, and we will introduce special cut rules describing how to handle them below.
The cut rules will result in larger regions $\VV$ to be removed from $\PP$ in iterations where the problematic object is an active cut arc.
We therefore say that the resulting algorithm is \emph{aggressive}.
The old types of problematic objects (convex vertices and convex arcs of radius less than $1$) are handled as by the basic algorithm, but only if there is no active cut arc in the same component of $\PP$.
The pseudocode of the resulting algorithm can be seen in Algorithm~\ref{roundAlg}.
One can think of the new cut rules as a way to shortcut the process of the basic algorithm (and variations of it as described in Section~\ref{sec:infinite}) so that instead of making an infinite sequence of cuts that converge to some arc in the limit, we directly make a cut defined by that limit arc.

We now describe when a new cut arc becomes an active cut arc.
Consider the situation where we have chosen an arc $\cut$ and thus removed a subset from a connected component $\KO$ of $\PP_i$ to obtain $\PP_{i+1}$.
The remaining part of $\KO$ consists of a number $t$ of components $\KO_1,\ldots,\KO_t$, and each of these $\KO_j$ will have one or more new cut arcs $\cut_1,\ldots,\cut_u$ which are contained in $\cut$.
If $u=1$, i.e., a component $\KO_j$ has only one new cut arc, and that arc $\cut_1$ is not perfect (which implies that one or both endpoints of $\cut_1$ is a convex vertex), then we define $\cut_1$ to be the \emph{active} cut arc of $\KO_j$.
Otherwise, no arc of $\KO_j$ is active.
Note in particular that at any time during the execution of the algorithm, each component of $\PP$ has at most one active cut arc.

\begin{algorithm}[h]
\LinesNumbered
\DontPrintSemicolon
\SetArgSty{}
\SetKwInput{Input}{Input}
\SetKwInput{Output}{Output}
\SetKw{Report}{report}
\SetKwIF{If}{ElseIf}{Else}{if}{}{else if}{else}{end if}
\SetKwFor{Foreach}{for each}{}{end for}
\SetKwFor{While}{while}{}{end while}
\While {a connected component $\KO$ of $\PP$ does not have bounded convex curvature} {\label{loop}
  Let $\se$ be the active cut arc of $\KO$ if there is one, and otherwise let $\se$ be a convex arc of radius less than $1$ or a convex vertex of $\KO$. \\
  Remove $\VV$ from $\KO$. \\
  For each resulting component $\KO'\subset\KO\setminus\VV$, if $\KO'$ has exactly one new cut arc $\cut'\subseteq\cut$ and $\cut'$ is not perfect, mark $\cut'$ as active. \\
}
\Return {$\PP$}
\caption{$\ttt{BoundedConvexCurvature}(\PP)$}
\label{roundAlg}
\end{algorithm}

In the following, we specify the cut rules when the problematic object is an active cut arc $\se$.
Figure~\ref{fig:cuttypesActive} shows an example of each type of cut.
Again, a cut will be specified by an arc $\cut$ with endpoints $\start(\cut)$ and $\terminal(\cut)$ on the boundary $\partial\PP$, as described in Section~\ref{sec:specifying}.
We define $\arcA$ to be the arc preceding $\se$ and $\arcB$ to be the arc succeeding $\se$ on $\partial\PP$.
The statement inside the square brackets [ ] specifies when the type of cut applies.

The intuition behind the rules is that in order to avoid infinite loops, we should prefer to make cuts that either completely removes $\arcA$ or $\arcB$ or such that the disk $\cutdisk(\cut)$ touches $\arcA$ and $\arcB$.

\begin{figure}
\centering
\includegraphics[page=20]{collectFigs.pdf}
\caption{An example of each type of cut when the problematic object $\se$ is an active cut arc.}
\label{fig:cuttypesActive}
\end{figure}

\begin{enumerate}
\item[3.]
[$\se$ is a cut arc with a convex endpoint $p\in\{\start(\se),\terminal(\se)\}$ and the other arc incident at $p$ is also a cut arc.]
Suppose without loss of generality that $p=\start(\se)$. 
We then define $\cut$ to be the unit radius arc from $\start(\arcA)$ to $\terminal(\se)$.

\item[4.]
[$\se$ is a cut arc with a convex endpoint, the other endpoint of $\sigma$ is tangential or convex, and each neighbour incident at a convex endpoint of $\sigma$ is not a cut arc.]
Consider the bisectors $\xi_{\arcA}\mydef\xi(\arcA,\se)$ and $\xi_{\arcB}\mydef \xi(\se,\arcB)$, and let $a$ be the end of $\xi_{\arcA}$ and $b$ be the end of $\xi_{\arcB}$.
\begin{enumerate}
\item[4.1.]
[$\arcA$ limits $\xi_{\arcA}$ and $\arcA$ is not a cut arc, or $\arcB$ limits $\xi_{\arcB}$ and $\arcB$ is not a cut arc.]
Without loss of generality, assume that $\arcA$ limits $\xi_{\arcA}$.
We then choose $\cut$ as the unit radius arc from $\start(\arcA)$ to $o\mydef \Pi_\se(\arcA,\se,a)$, where $a$ is the end of $\xi_{\arcA}$.

\begin{convention}\label{conv:double}
Let $\KO$ be the component where $o$ appears on the boundary after the cut, and let $\cut'$ be the part of $\cut$ that is an arc of $\KO$ incident at $o$.
Then $o$ is a convex vertex shared by the two cut arcs $\cut'$ and $\se'\mydef\se[o,\terminal(\se)]$.
If $\cut'$ is the only new cut arc of its component, then $\cut'$ will be an active cut arc.
Hence, the next cut in $\KO$ must be of type 3.
However, if the other neighbour of $\cut'$ is also a cut arc $D$, it is not specified if we should cut from $\start(D)$ to $\terminal(\cut')$ or from $\start(\cut')$ to $\terminal(\se')$.
For technical reasons (to be used when bounding the number iterations made by the algorithm in the proof of Lemma~\ref{linearIts}) we make the convention that we always choose the last of these two options, so that after the next cut, both $\cut'$ and $\se$ will have disappeared from $\partial\PP$.
\end{convention}

\item[4.2.]
[Otherwise], there exists a disk $C$ of radius at most $1$ that touches $\arcA$, $\se$, and $\arcB$.
Let $x_0$ be the center of $C$.
We then consider the bisector $\xi\mydef\xi(\arcA,\arcB)$ and follow $\xi$ from $x_0$ until we get to a point $x$ where $C_x$ has radius $1$ or we get to the end of $\xi$, so we have two types~4.2.1 and 4.2.2 as for type 2.
\end{enumerate}

\item[5.]
[$\se$ is a cut arc with a convex endpoint $p\in\{\start(\se),\terminal(\se)\}$, the other arc incident at $p$ is not a cut arc, and the other endpoint of $\se$ is concave.]
Without loss of generality, assume that $p=\start(\se)$.
We have two subtypes.
\begin{enumerate}
\item[5.1.]
[There exists a disk $C$ of radius $1$ touching $\arcA$ at a point $a$ and containing $\terminal(\se)$.]
If there are more, we choose $C$ with the contact point $a$ closest to $\terminal(\arcA)$.
We then define $\cut$ as the unit radius arc from $a$ to $\terminal(\se)$.

\item[5.2.]
[Otherwise], we define $\cut$ to be the unit radius arc from $\start(\arcA)$ to $\terminal(\se)$.
\end{enumerate}

\end{enumerate}

This finishes the description of our aggressive algorithm.

\subsection{Basic properties of aggressive cut rules}

We now prove that the cut rules for handling active cut arcs have the same properties as the ones described in Section~\ref{sec:basicSimple}.
We therefore repeat the lemmas and give proofs for the new cut rules when needed.

\simplecurvelemma*

\begin{figure}
\centering
\includegraphics[page=14]{collectFigs.pdf}
\caption{Figures for the proof of Lemma~\ref{lem:simplecurve}.
Left: Fact~\ref{fact:fact} says that the unit arcs from $x$ to $y$ and $x$ to $z$ only intersect at $x$.
Right: A special case of a cut of type 4.1.
}
\label{fig:factetc}
\end{figure}

\begin{proof}[Proof for cut rules 3--5.]
We use the following fact which is easy to check; see Figure~\ref{fig:factetc} (left) for an illustration.

\begin{fact}\label{fact:fact}
Let $x,y,z$ be three points such that the distance from $x$ to each of $y$ and $z$ is at most $2$.
Then the unit radius arcs from $x$ to $y$ and from $x$ to $z$ have only the point $x$ in common, unless one arc is contained in the other.
\end{fact}

We now divide into the type of cut, as follows.
\begin{enumerate}
\item[3.]
We need to prove that 
$\cut\cap\arcA=\{\start(\arcA)\}$ and $\cut\cap \se=\{\terminal(\se)\}$.
Let $D_{\arcA}$ be the unit disk with $\arcA\subset\partial D_{\arcA}$ and define $D_{\se}$ similarly.
We first observe that $\terminal(\se)\in D_{\arcA}$ or $\start(\arcA)\in D_\se$: if both of these properties are violated, it is easy to see that $\arcA$ and $\se$ have two intersection points, which contradicts that the boundary $\partial \PP$ is a simple, closed curve.
So suppose without loss of generality that $\terminal(\se)\in D_\arcA$.
It then follows that the distance between $\start(\arcA)$ and $\terminal(\se)$ is at most $2$.
Since we know that $\arcA$ and $\se$ both span angles of at most $\pi$, we also have that $\arcA$ is the unit radius arc from $\start(\arcA)$ to $\terminal(\arcA)=\start(\se)$ and that $\se$ is that from $\terminal(\arcA)=\start(\se)$ to $\terminal(\se)$.
It now follows from Fact~\ref{fact:fact} that $\arcA$ and $\cut$ have only the point $\start(\arcA)$ in common and $\se$ and $\cut$ have only the point $\terminal(\se)$ in common, so the statement follows.

\item[4.]

Note that it follows for cuts of this type that if $\arcA$ (resp.~$\arcB$) is a cut arc, then $\se$ and $\arcA$ (resp.~$\se$ and $\arcB$) are tangential.
If the shared vertex is convex, we should make a cut of type 3 instead.

\begin{enumerate}
\item[4.1]
We consider without loss of generality the case that $\arcA$ limits $\xi_{\arcA}$ and $\arcA$ is not a cut arc.
We choose $\cut$ in the disk $C_a$ touching $\arcA$ and $\se$, where the center $a$ is the end of $\xi_{\arcA}$.
Suppose that $C_a\cap \arcA$ consists of more than one point.
As for cuts of type 2, we conclude that $\arcA$ and $\se$ are tangential at $\terminal(\arcA)=\start(\se)$ and that $\arcA\subset\partial C_a$.
Since $\arcA$ is not a cut arc, we get that $\arcA$ has radius less than $1$ and that $C_a$ touches $\se$ at $\start(\se)$, as shown in Figure~\ref{fig:factetc} (right).
The arc $\cut$ is therefore the unit radius arc from $\start(\arcA)$ to $\terminal(\arcA)$.
The cut is similar to one of type 1, so the result follows.

Suppose now that $C_a\cap \arcA$ consists of one point and that $C_a\cap\se$ consists of more than one point.
Again, we conclude that $\se$ and $\arcA$ are tangential at $\start(\se)$ and that $\se\subset\partial C_a$.
Now, since $\arcA$ limits $\xi_{\arcA}$, we conclude that $\arcA$ is also a cut arc, but that contradicts our assumption.

On the other hand, if $C_a\cap \arcA$ and $C_a\cap\se$ both consist of single points, then it follows directly that $\cut\cup \partial\PP(\start(\cut),\terminal(\cut))$ is a simple, closed curve, as $\cut\subset C_a$.

\item[4.2]
We first prove that $C_x \cap\arcA$ and $C_x \cap\arcB$ consist of single points.
For instance, suppose that $C_x\cap\arcA$ has more points.
Then $\arcA$ must be an arc on $C_x$, and since $C_x$ is a unit disk, $\arcA$ is a cut arc.
But then $\arcA$ and $\se$ are two cut arcs with a common vertex $\start(\se)$ which must be tangential.
We conclude that $\terminal(\se)$ must be convex in order for the cut to be of type 4, but then $\arcB$ enters the interior of $C_x$ at $\terminal(\se)=\start(\arcB)$, which contradicts that $C_x$ touches $\arcB$.
Hence, $\cut \cap\arcA$ and $\cut \cap\arcB$ consist of single points.

\begin{figure}
\centering
\includegraphics[page=23]{collectFigs.pdf}
\caption{Figures for the proof of Lemma~\ref{lem:simplecurve}.
Left: For a cut of type 4.2, it is not possible that the arcs $\sigma$ and $\cut_{y'}$ are tangential, since that would separate $\terminal(\sigma)$ from $\terminal(\cut_{y'})$ so that $\arcB$ cannot connect them.
Right: For a cut of type 5.2, it leads to a contradiction if $\cut$ and $\arcA$ have an extra intersection point $p$.}
\label{fig:factetc2}
\end{figure}

It remains to be verified that $\cut$ and $\se$ are disjoint.
Recall that we defined $x_0$ to be a point such that the disk $C_{x_0}$ touches $\arcA$, $\arcB$, and $\se$.
We then follow $\xi=\xi(\arcA,\arcB)$ from $x_0$ until we get to a point $x$ where $C_x$ has radius $1$ or we get to the end of $\xi$.
For any point $y\in\xi[x_0,x]$, we define $\cut_y$ to be the unit radius arc from $\Pi_{\arcA}(y)$ to $\Pi_{\arcB}(y)$.
If $\cut$ and $\se$ are not disjoint, we may consider the first point $y'\in \xi[x_0,x]$ after $x_0$ where $\cut_{y'}$ and $\se$ intersect.
This may happen for two different reasons: (i) $\cut_{y'}$ contains an inner point of $\se$ or (ii) $\cut_{y'}$ contains an endpoint of $\se$.
Consider first case (i).
It is not possible that an intersection point in $\cut_{y'}\cap \se$ is an endpoint of $\cut_{y'}$, because then $\se$ would intersect $\arcA$ or $\arcB$ at points other than $\terminal(\arcA)$ and $\start(\arcB)$.
Hence, $\cut_{y'}$ and $\se$ are tangential and touch each other at a unique point $c$ which is an inner point of both arcs; see Figure~\ref{fig:factetc2} (left).
It is now easy to verify that the simple, closed curve $\gamma=\cut_{y'}[\start(\cut_{y'}),c]\cup\se[c,\start(\se)]\cup\arcA[\start(\cut_{y'}),\start(\se)]$ separates $\start(\arcB)=\terminal(\se)$ from $\terminal(\cut_{y'})\in\arcB$.
Hence, $\arcB$ has to intersect $\gamma$, which is impossible.
We now consider case (ii).
If $\cut_{y'}$ and $\se$ intersect at an endpoint of $\cut_{y'}$, we get that $\arcA$ or $\arcB$ self-intersect or intersect each other, which cannot happen.
But it also cannot happen that $\cut_{y'}$ and $\se$ intersect at an inner point of $\cut_{y'}$, since $\cut_{y'}$ is contained in the disk $C_{y'}$ which intersects $\arcA$ and $\arcB$ only in the points $\start(\cut_{y'})$ and $\terminal(\cut_{y'})$.
\end{enumerate}

\item[5.]
In both types 5.1 and 5.2, we get from Fact~\ref{fact:fact} that $\cut$ and $\se$ intersect only at $\terminal(\se)$.
\begin{enumerate}
\item[5.1]
Since $\cut$ and $\arcA$ are tangential, they can have only one intersection point, and the statement follows.

\item[5.2]
Suppose for contradiction that $\cut$ and $\arcA$ in addition to the point $\start(\cut)=\start(\arcA)$ also intersect at a point $p$; see Figure~\ref{fig:factetc2} (right).
Let $C$ be the unit disk containing $\cut$.
We now consider rotating $C$ clockwise so that $\partial C$ always contains $\terminal(\se)$.
Then the intersection points $\arcA\cap \partial C$, which are initially $\start(\arcA)$ and $p$, will move towards each other.
Eventually, there will be a single intersection point somewhere in between the two intersection points $\arcA\cap\cut$, and $C$ will be a unit disk touching $\arcA$ and with $\terminal(\sigma)\in\partial C$, thus fulfilling the requirements for a cut of type 5.1.
Hence, the cut should not have been of type 5.2, but of type 5.1.
\qedhere
\end{enumerate}
\end{enumerate}
\end{proof}

We now obtain Lemma~\ref{lem:unitpropSimple} for our new cut rules.
The argument is similar to the one in the proof for cuts of type 2.

\unitdisklemma*

\subsection{Correctness of the aggressive algorithm}

We now obtain Lemma~\ref{lemma:algCurve} for Algorithm~\ref{roundAlg}, by an unchanged proof.

\lemmaalgcurve*

We can finally state the correctness of Algorithm~\ref{roundAlg}, which again follows by an identical proof.

\thmcorrect*

\subsection{Linear bound on the number of iterations}
\label{sec:linear}

In order to bound the number of iterations,  we first bound the number of some special types of cut arcs and cuts, which are then used to bound the remaining ones.

\begin{lemma}\label{lem:perfCutArcs}
At most $O(n)$ perfect cut arcs are created by Algorithm~\ref{roundAlg}.
\end{lemma}

\begin{proof}
We show that each time we make a perfect cut arc, it corresponds to drawing an edge in a directed plane graph $G$ with $2n$ vertices.
In this graph, there can be multiple edges from one vertex to another, but whenever that is the case, the region enclosed by two neighbouring edges will contain another vertex.
It hence follows from a variant of Euler's formula that there can be at most $O(n)$ edges in total.

We first observe that when making a perfect cut arc $\cut'$, each endpoint of $\cut'$ is either an original vertex of $\PP_1$ or a point on an original arc of $\PP_1$:
Otherwise, an endpoint $p$ of $\cut'$ is on another cut arc made in a previous iteration.
But then $p$ is a convex vertex, so $\cut'$ is not perfect.

The vertices of our graph $G$ are the original vertices of $\PP_1$ and the original arcs of $\PP_1$.
Here, each arc excludes the endpoints, such that the vertices of $G$ are pairwise disjoint.

Now suppose that we make a perfect cut arc $\cut'$ by cutting along an arc $\cut\supset\cut'$ in order to handle a problematic object.
We then draw an edge from $\start(\cut')$ to $\terminal(\cut')$ contained in $\VV=\VV(\cut)$.
Since the vertices of $G$ are pairwise disjoint, this uniquely defines the vertices in $G$ that we connect.
Because we draw the edge in the region $\VV$ that we remove, the edges from different iterations cannot cross and the result is a plane graph.
Suppose that we draw multiple edges $e_1,\ldots,e_k$ from one vertex to another in $G$ that appear in this cyclic order around one of the vertices.
We now prove that for each $j\in\{1,\ldots,k-1\}$, the region $F_j$ enclosed by $e_j$ and $e_{j+1}$ contains a hole of $\PP_1$.
This translates to saying that $F_j$ contains a vertex of $G$.
Undirected plane graphs with this property and $n$ vertices are known to have at most $3n-5$ edges~\cite{abrahamsen2020tiling}, and our graph $G$ has $2n$ vertices and is directed.
We can partition the edges into two sets so that in each set, there are no anti-parallel edges.
Hence, there are at most $6n-5$ edges in each set, so $G$ has at most $12n-10$ edges.

\begin{figure}
\centering
\includegraphics[page=3]{collectFigs.pdf}
\caption{Situation in the proof of Lemma~\ref{lem:nperfectcutarcs}.
The region $F$ is gray (light and dark), and $M\subset F$ is dark gray.
It is impossible that the algorithm removes a subset $\VV_k$ of $M$.}
\label{fig:nottwoarcs}
\end{figure}

For any pair of vertices $p,q$ of $\PP_1$, there is at most one unit radius arc from $p$ to $q$, so there can be made at most one perfect cut arc from $p$ to $q$, and it remains to bound the number of perfect cut arcs with an endpoint on an inner point of an arc of $\PP_1$.
Suppose that we draw two edges from $\arcA$ to $\arcB$, where $\arcA$ and $\arcB$ are arcs of $\PP_1$; see Figure~\ref{fig:nottwoarcs}.
The case where we draw an edge between a vertex and an edge follows from a similar reasoning.
Say that in iterations $i$ and $j$, where $i<j$, we introduce perfect cut arcs $\cut_i$ and $\cut_j$, respectively, connecting $\arcA$ and $\arcB$.
Let $e_i$ and $e_j$ be the edges we draw in $G$ corresponding to $\cut_i$ and $\cut_j$.
Let $F$ be the region bounded by the simple, closed curve $\arcA[\start(\cut_j),\start(\cut_i)]\cup e_i\cup\arcB[\terminal(\cut_i),\terminal(\cut_j)]\cup e_j$, which corresponds to the region enclosed by $e_i$ and $e_j$ in the graph $G$.

Suppose for contradiction that $F$ does not contain a hole of $\PP_1$.
Then, since $e_j\subset\PP_j\subset\PP_i$, it follows that $F\subset\PP_i$.
As $\cut_i\subset F\subset\PP_i$, we then have that $\cut_i$ becomes one cut arc of $\PP_{i+1}$.

Define $M\subset F$ to be the region bounded by the simple, closed curve $\partial\PP_{i+1}[\start(\cut_j),\terminal(\cut_j)]\cup e_j$, consisting of $\cut_i$, $e_j$, and parts of $\arcA$ and $\arcB$.
There is no convex vertex on the boundary $\partial\PP_{i+1}[\start(\cut_j),\terminal(\cut_j)]$ since the endpoints of the arc $\cut_i$ are tangential.
Since the algorithm later cuts along $\cut_j$, a convex vertex must appear in $M$ in some iteration $k$ in between the two cuts, i.e., $i<k<j$.
The cut in iteration $k$ removes a region $\VV_k$ and since it creates a convex vertex in $M$, we must have $\VV_k\subset M$.
But the only boundary of $\PP_{i+1}$ in $M$ is $\partial\PP_{i+1}[\start(\cut_j),\terminal(\cut_j)]$, which has no convex vertex, so the algorithm would not remove such a region $\VV_k$, which is a contradiction.
\end{proof}

We now study cuts of type 4.2.1 and 5.1.
These are special in that they are the only cuts where we handle a cut arc $\sigma$ and parts of both of the neighbours $\arcA$ and $\arcB$ of $\se$ remain on $\partial\PP$ after the cut.
In all other cuts handling a cut arc $\sigma$, one of the neighbours of $\sigma$ is completely removed.

\begin{lemma}\label{lem:nperfectcutarcs}
Algorithm~\ref{roundAlg} makes at most $O(n)$ cuts of types 4.2.1 and 5.1.
\end{lemma}

\begin{proof}
The argument is similar to that used in the proof of Lemma~\ref{lem:perfCutArcs}:
We define a plane graph $G$, where the vertices of $G$ are the vertices and edges of $\PP_1$.

We first observe that when making a cut of type 4.2.1, the neighbours $\arcA$ and $\arcB$ of $\sigma$ must both be original arcs of $\PP_1$:
Otherwise one of them, say $\arcA$, is a cut arc, and the unit disk $\cutdisk=\cutdisk(\cut)$ touches $\arcA$, so $\arcA\subset \partial\cutdisk$.
The vertex $\start(\se)$ must be tangential (otherwise we should make a cut of type 3), and therefore we also have $\se\subset\partial\cutdisk$.
Hence $\terminal(\se)$ is convex (otherwise $\se$ would be perfect), so $\cutdisk$ does not touch $\arcB$, which is a contradiction.

For a cut of type 5.1, assume without loss of generality that $\cutdisk$ touches $\arcA$ and that $\start(\arcB)\in \partial \cutdisk$.
Then $\arcA$ must be an arc of $\PP_1$, by a similar reasoning as for type 4.2.1.
Furthermore, $\start(\arcB)$ is a concave vertex by definition of the cut type, so the vertex must be an original vertex of $\PP_1$, since it is easy to check that all new vertices made by the algorithm are either tangential or convex.
Therefore the objects we have included as vertices in our graph $G$ are sufficient.
We can then proceed as in the proof of Lemma~\ref{lem:perfCutArcs}.
\end{proof}

\begin{figure}
\centering
\includegraphics[page=21]{collectFigs.pdf}
\caption{Situation in the proof of Lemma~\ref{lemma:holearcs}.
The shown cut splits a component into four pieces, eliminates four holes, and creates seven hole cut arcs.}
\label{fig:lemma:holearcs}
\end{figure}

Consider a component $\KO$ of $\PP$ with new cut arcs $\cut_1,\ldots,\cut_u$.
If $u>1$, we denote these arcs as \emph{hole cut arcs}, since their creation corresponds to the elimination of $u-1$ holes in $\PP$, as explained in the proof of the following lemma.

\begin{lemma}\label{lemma:holearcs}
Less than $2n$ hole cut arcs are created by Algorithm~\ref{roundAlg}.
\end{lemma}

\begin{proof}
Consider the arc $\cut$ that defines the cut to be made in $\PP_i$ in order to obtain $\PP_{i+1}$.
Suppose that the cut creates hole cut arcs $\cut_1,\ldots,\cut_t$ of $\PP_{i+1}$ that appear in this cyclic order along $\cut$; see Figure~\ref{fig:lemma:holearcs}.
We choose a point $p\in\VV=\VV(\cut)$, and in each connected component $\KO$ of $\PP_{i+1}$, we choose a point $q_{\KO}\in\KO$.
For each $j\in \{1,\ldots,t\}$, we now choose an open curve $\gamma_j$ from $p$ to $q_\KO$, where $\KO$ is the component of $\cut_j$.
We choose a curve $\gamma_j$ that consists of one interval contained in $\VV$ and the other in $\KO$, so that $\gamma_j$ leaves $\VV$ through a point on $\cut_j$.
Furthermore, these curves $\gamma_1,\ldots,\gamma_t$ can be chosen to be pairwise disjoint except for the endpoints.
For a pair $\cut_j,\cut_k$ of hole cut arcs that are consecutive on the same component of $\PP_{i+1}$, the closed curve $\gamma_j\cup \gamma_k$ enclose a hole $H_j$ where $\terminal(\cut_j)\in\partial H_j$, which we associate to the arc $\cut_j$.
We now observe that no other hole cut arc can be associated to the same hole $H_j$:
The curve $\gamma_j\cup \gamma_k$ separates $H_j$ from all hole cut arcs in the exterior of $\gamma_j\cup \gamma_k$, and a hole cut arc $\cut_l$ in the interior of $\gamma_j \cup \gamma_k$ can only be associated to a hole $H_l$ which is separated from $H_j$ by another curve $\gamma_l \cup \gamma_m$ contained in the interior of $\gamma_j \cup \gamma_k$.
Each hole $H_j$ is merged with the exterior of $\PP$ due to the cut, so the number of holes decreases by at least $t/2$.
Since we start with less than $n$ holes, it follows that we create less than $2n$ hole cut arcs in total.
\end{proof}

\begin{lemma}\label{linearIts}
Algorithm~\ref{roundAlg} performs $O(n)$ iterations and makes in total $O(n)$ cut arcs.
\end{lemma}

\begin{proof}
We make an amortized analysis where we assign credits to the arcs of $\PP$ which are used to ``pay'' for later iterations.
All in all, we will create $O(n)$ credits, and each credit can be used to pay for one iteration, so the statement follows.

\begin{figure}
\centering
\includegraphics[page=15]{collectFigs.pdf}
\caption{Left: The fat arcs are the new cut arcs $\cut_1,\cut_2,\cut_3$. The arc $\cut_3$ is a double cut arc and is thus the union of two arcs.
Right: The arc $\se$ is an active cut arcs and the arcs have credits according to the invariant.}
\label{fig:doublecutarc}
\end{figure}

Before we describe the actual accounting scheme, we explain the concept of a \emph{double cut arc}, which allows us to express the invariant of our accounting scheme in a much simpler form:
Suppose that we have picked an active cut arc $\se$.
Then $\se$ is completely removed from $\partial\PP$ except if the cut is of type 4.1.
In a cut of type 4.1, a part of $\se$ will remain and be adjacent to one of the new cut arcs created by the cut.
To be precise, let the new cut arcs created be $\cut_1,\ldots,\cut_t$, and suppose that $\se'\subset \se$ is the part of $\se$ that remains; see Figure~\ref{fig:doublecutarc} (left).
Then $\se'$ is a neighbour of $\cut_1$ or $\cut_t$.
Without loss of generality, consider the second case.
We then consider $\se'\cup\cut_t$ to be a single cut arc, which we call a \emph{double cut arc}.
By slight abuse of notation, we shall from now on (but only in this proof) denote this double cut arc as $\cut_t$.
By Convention~\ref{conv:double}, we know that if the double cut arc $\cut_t$ is active, then it will disappear by the following cut in its component due to a cut of type~3.

We can now express the desired invariant; see Figure~\ref{fig:doublecutarc} (right).
Initially, we put $4$ credits on each arc of the original input $\PP_1$.
We will keep the following invariant. 

\textbf{Invariant:}
\emph{
In each connected component $\KO$ of $\PP$, the following holds:
If $\KO$ has an active cut arc $\se$, then $\se$ has at least $1$ credit.
All arcs of $\KO$ that are not active or neighbours of an active cut arc have at least $4$ credits.
}

Let $\se$ be the element 
that the algorithm is handling in one iteration.
If $\se$ is an active cut arc, then it has at least $1$ credit by the invariant, and this is used to pay for the present iteration.
Otherwise, all arcs in the component of $\se$ have $4$ credits and $\se$ is a convex vertex or a convex arc with radius less than $1$.
If $\se$ is a convex vertex, we use $1$ credit from one of the incident arcs to pay for the iteration.
If $\se$ is a convex arc of radius less than $1$, then $\se$ has $4$ credits and we use $1$ to pay for the iteration.
We now show that by paying for the iterations in this way, we are able to maintain the invariant.

\begin{figure}
\centering
\includegraphics[page=10]{collectFigs.pdf}
\caption{Left: A cut creates three new cut arcs $\cut_1,\cut_2,\cut_3$ and their neighbours $\arcA_j,\arcB_j$ are shown. (Not all arcs of $\PP$ are shown.)
Right: An example where $\cut_2$ is a double cut arc and $\bar\arcB_1=\bar\arcA_2=\bar\arcB_2$.}
\label{fig:arcAB}
\end{figure}

Consider the arc $\cut$ that defines the cut to be made in order to handle the object $\se$.
Suppose that the cut generates $t$ cut arcs $\cut_1,\ldots,\cut_t$, all contained in $\cut$, which are the new cut arcs of $\PP_{i+1}$.
Here, $\cut_1$ or $\cut_t$ may be a double cut arc (which is the case if the cut is of type 4.1).
Let $\arcA_j$ and $\arcB_j$ be the arcs of $\PP_{i+1}$ preceding and succeeding $\cut_j$; see Figure~\ref{fig:arcAB} (left).
We denote the arcs $\arcA_1,\arcB_1,\ldots,\arcA_t,\arcB_t$ as the \emph{neighbour arcs} of $\PP_{i+1}$.

Some of the new cut arcs $\cut_1,\ldots,\cut_t$ will be active in their component.
Recall that this is the case for an arc $\cut_j$ if $\cut_j$ is the only new cut arc in its component and $\cut_j$ is not perfect.
We need to ensure that we can assign $1$ credit to the active cut arcs and $4$ to all the others and their neighbours.

If $\cut_j$ is a new perfect cut arc, we can create $12$ new credits so that we can assign $4$ credits to $\cut_j$ and the neighbour arcs $\arcA_j$ and $\arcB_j$.
By Lemma~\ref{lem:perfCutArcs}, this results in the creation of at most $O(n)$ credits in total.
In the following, we prove that we can also assign enough credits to the new cut arcs that are not perfect and their neighbours.

The arcs $\arcA_j$ and $\arcB_j$ stem from arcs $\bar{\arcA_j}$ and $\bar{\arcB_j}$ of $\PP_i$, respectively, which contain $\arcA_j$ and $\arcB_j$, but are generally longer.
We denote the arcs $\bar\arcA_1,\bar\arcB_1,\ldots,\bar\arcA_t,\bar\arcB_t$ as the \emph{original neighbour arcs}.
It is possible that different neighbour arcs of $\PP_{i+1}$ stem from the same original neighbour arc of $\PP_i$.
For instance, we can have $\bar\arcA_j=\bar\arcB_j$ or $\bar\arcB_j=\bar\arcA_{j+1}$.
Furthermore, it may be the case that the outermost original neighbour arcs $\bar\arcA_1$ and $\bar\arcB_t$ have no credits because they are neighbours of the active cut arc $\se$ in $\PP_i$.
As we will see, enough of the original neighbour arcs will have $4$ credits so that they can be distributed in a way that satisfies the invariant.

Let us consider the event that three of the neighbour arcs stem from $\bar\arcB_t$; see Figure~\ref{fig:arcAB} (right) for an example where this happens.
We claim that this is possible only if 
$\cut_t$ is a double cut arc.
If 
$\cut_t$ is not double, then $\bar\arcB_t$ intersects $\cut$ three times, which is impossible.
Similarly, at most three neighbour arcs can stem from $\bar\arcA_1$.
We also see that it is impossible that four neighbour arcs stem from $\bar\arcA_1$ or $\bar\arcB_t$, since that would imply that one of these arcs had three intersection points with $\cut$.
We conclude that at most six of the neighbour arcs stem from $\bar\arcA_1$ or $\bar\arcB_t$.
Similarly, consider an original neighbour arc $\bar\arcA_j$ or $\bar\arcB_j$ which is not $\bar\arcA_1$ or $\bar\arcB_t$.
It is possible that two neighbour arcs stem from such an arc, but not more than that.

With these observations in mind, we now prove that the scheme for redistributing credits works.
We first consider the components of $\PP_{i+1}$ that have more than one new cut arc contained in $\cut$. 
If $k\geq 2$ of the arcs $\cut_1,\ldots,\cut_t$ are on the same component $\KO$ of $\PP_{i+1}$, then these are hole cut arcs.
We create $4$ credits for each of the $k$ new cut arcs of $\KO$ and their neighbours, and the invariant is then satisfied for $\KO$.
By Lemma~\ref{lemma:holearcs}, this results in the generation of at most $O(n)$ credits in total.

We now define $t'\leq t$ to be the number of components created that have exactly one active cut arc contained in $\cut$, and we need to verify that we can distribute the credits so that these $t'$ new cut arcs get $1$ credit each.
We divide into two cases depending on the number $t'$ of these components:
\begin{itemize}
\item[$t'\leq 3$]
If $\bar\arcA_1$ or $\bar\arcB_t$ has $4$ credits, the $t'$ arcs can get a credit each and the invariant holds.
When $\bar\arcA_1$ and $\bar\arcB_t$ have less, the object $\se$ must have been an active cut arc, and $\bar\arcA_1$ and $\bar\arcB_t$ must be the neighbours of $\se$ on $\partial\PP_i$.
Furthermore, $\cut$ is chosen according to a cut of type 4.2.1 and 5.1, since for cuts of all other types, one of the neighbours of $\se$ is completely removed, and then one of $\bar\arcA_1$ and $\bar\arcB_t$ would have $4$ credits.
By Lemma~\ref{lem:nperfectcutarcs}, there are only $O(n)$ cuts of type 4.2.1 and 5.1, so we can afford to generate $O(1)$ new credits every time to be placed on the new arcs.

\item[$t'\geq 4$]
Since at most six of the neighbour arcs can stem from $\bar\arcA_1$ or $\bar\arcB_t$, there are at least $2(t-3)$ that do not.
These stem from at least $t-3$ unique original neighbour arcs of $\PP_i$ that have $4$ credits each, so we have $4(t-3)\geq t\geq t'$ credits available.
Hence, there are enough credits for the $t'$ new cut arcs to get $1$ credit each.
\end{itemize}

Since each new cut arc gets at least $1$ credit and we generate at most $O(n)$ credits, it follows that the algorithm makes in total $O(n)$ cut arcs.
\end{proof}

We have now proved the following theorem.

\begin{theorem}\label{LTThm}
Given a curvilinear region $\PP_1$ bounded by $n$ line segments and circular arcs,
there is a unique maximum set $\QQ\subseteq\PP_1$
of bounded convex curvature
which contains every set $\QQ'\subseteq\PP_1$
of bounded convex curvature.
At any time during the execution of Algorithm~\ref{roundAlg}, the boundary
$\partial\PP$ consists of $O(n)$ line segments and circular arcs, and
the algorithm returns the result $\QQ$ after $O(n)$ iterations.
\end{theorem}

\section{Implementation}\label{anImplementation}

Here we describe two implementations of the algorithm.
The first one runs in $O(n^2)$ time and works for general curvilinear regions in the standard DCEL (doubly connected edge list) representation, with no additional data structures required.
The second runs in $O(n\log n)$ time and works under the assumption that the input $\PP_1$ is a simply-connected curvilinear region.
This assumption is needed to make use of efficient data structures for circular ray shooting and fully dynamic orthogonal range searching.
As is often the case for geometric algorithms, it may turn out challenging to program an actual implementation of these algorithms in practice.
For instance, circular arcs intersecting in non-generic ways can lead to robustness issues when using floating-point arithmetic.
We ignore these issues and refer the reader to the work by Devillers, Fronville, Mourrain, and Teillaud~\cite{DEVILLERS2002119}, who described a method to design exact geometric predicates in algorithms dealing with curved objects such as circular arcs.

\subsection{General curvilinear regions}\label{anImplementation:general}

For each connected component of $\PP$, the algorithm stores a pointer to the active cut arc, if there is one.
If there is a component with an active cut arc, the algorithm handles that arc.
Otherwise, the algorithm traverses the boundary of $\PP$ to check if there is a convex vertex or a convex arc with radius less than $1$.
If there is none, $\PP$ is returned since it is our maximum subset of bounded convex curvature.

Suppose that a problematic object has been chosen, and let $\cut$ be the arc specifying a cut to be performed in $\PP$.
Algorithm~\ref{performCut} traverses the boundary of the region $\VV=\VV(\cut)$, that we must remove, from $\start(\cut)$ clockwise to $\terminal(\cut)$ (since portions of $\cut$ are concave arcs of $\VV$, this corresponds to traversing $\cut$ in counterclockwise direction).
Note that the remaining part of $\partial\VV$ is $\partial\VV[\start(\cut),\terminal(\cut)]=\partial\PP[\start(\cut),\terminal(\cut)]$, so the traversed part contains all the new cut arcs that should be created.
The algorithm changes the boundary $\partial\PP$ accordingly by splicing in the new cut arcs $\cut[a',b]\subset\cut$.

\begin{algorithm}[h]
\LinesNumbered
\DontPrintSemicolon
\SetArgSty{}
\SetKwInput{Input}{Input}
\SetKwInput{Output}{Output}
\SetKw{Report}{report}
\SetKwIF{If}{ElseIf}{Else}{if}{}{else if}{else}{end if}
\SetKwFor{Foreach}{for each}{}{end for}
\SetKwFor{While}{while}{}{end while}
$a\mydef \ttt{TraverseP}(\cut,\start(\cut))$.\; \label{alg1:line1}
\Repeat {$a=\terminal(\cut)$}
{
$a'\mydef a$.\;
$b\mydef \ttt{TraverseC}(\cut,a)$.\;
$a\mydef\ttt{TraverseP}(\cut,b)$.\;
If $a'$ (resp.~$b$) is not a vertex of $\PP$, then create one by splitting the arc containing $a'$ (resp.~$b$) into two. \;
Change $\partial\PP$ by setting the arc succeeding $a'$ and preceding $b$ to $\cut[a',b]$.\;
}
\caption{$\ttt{PerformCut}(\cut)$}
\label{performCut}
\end{algorithm}

When we call the subroutine $\ttt{TraverseP}(\cut,b)$, we assume that $b\in\cut\cap\VV$.
The subroutine traverses $\partial\PP$ from $b$ in clockwise direction until we reach a point $a$ such that (i) $\partial\PP$ exits $\VV$ at $a$, or (ii) $a=\terminal(\cut)$.
The point $a$ is then returned.
It will always be the case that $a\in\cut\cap\VV$, and it may be that $a=b$.
Case (ii) is just used to make sure that we stop when all new cut arcs have been constructed.

When we call the subroutine $\ttt{TraverseC}(\cut,a)$, it is assumed that $\partial\PP$ leaves $\VV$ at $a$ when traversed in clockwise direction.
The subroutine then follows $\cut$ in counterclockwise direction from $a$ through the interior of $\PP$ until we reach a point $b\in\partial\PP$, which is returned.
It then holds that $b\in\cut\cap\VV$, and it will always be the case that $a\neq b$.
The call to $\ttt{TraverseP}$ in line~\ref{alg1:line1} is needed because it can happen that a portion of $\partial P$ after $\start(\cut)$ (in clockwise direction) is contained in $\VV$, and therefore we cannot call $\ttt{TraverseC}(\cut,\start(\cut))$ directly.

The point $\ttt{TraverseC}(\cut,a)$ is the first intersection point between $\cut$ and $\partial\PP$ when following $\cut$ from
$a$.
This is found simply by checking all arcs of $\PP$.

\begin{theorem}\label{finalThmGen}
Algorithm~\ref{roundAlg} can be implemented so that it runs in time $O(n^2)$ and uses $O(n)$ space when the input is a general curvilinear polygon.
\end{theorem}

\begin{proof}
Each time the algorithm chooses a problematic object to be handled, it traverses $\partial\PP$.
By Theorem~\ref{LTThm}, this happens $O(n)$ times and each traversal takes $O(n)$ time, so this takes in total $O(n^2)$ time.
It then remains to bound the time used on $\ttt{PerformCut}$.

The running time of executing $\ttt{PerformCut}$ is clearly dominated by the time used at $\ttt{TraverseP}$ and $\ttt{TraverseC}$.
By Lemma~\ref{linearIts}, we make $O(n)$ calls to $\ttt{TraverseC}$, and by Theorem~\ref{LTThm}, each takes $O(n)$ since that is the number of arcs of $\PP$ at any time.
We therefore use $O(n^2)$ on $\ttt{TraverseC}$ in total.

The portion of $\partial\PP$ we traverse when calling
$\ttt{TraverseP}(\cut,b)$ is removed from $\PP$ and hence never traversed again.
Therefore, the time used on $\ttt{TraverseP}$ is bounded by the number of
vertices occuring on $\partial\PP$ during the execution of the algorithm.
By Lemma~\ref{linearIts}, this is $O(n)$ time in total.
\end{proof}

\subsection{Simply-connected curvilinear regions}\label{anImplementation:simply}

\begin{figure}
\centering
\includegraphics[page=1]{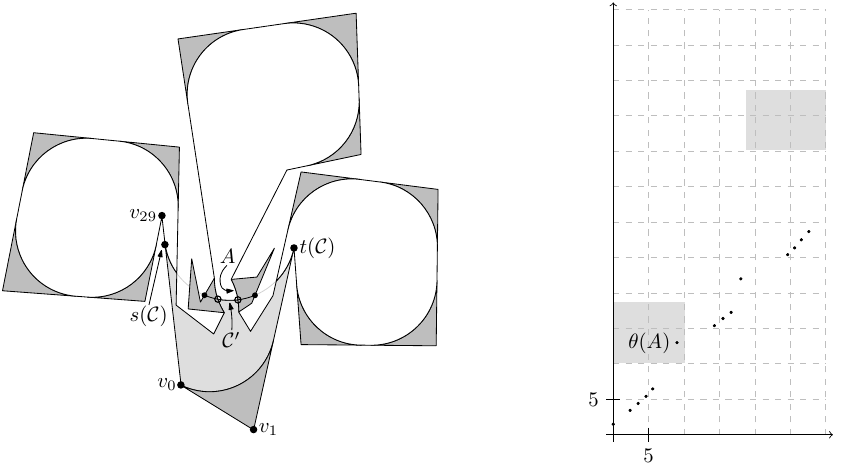}
\caption{Left: A cut of type 4.2.2 splits a component into three.
The middle component gets an improper active cut arc $\cut'$ with both neighbours also improper.
Right: The points $\theta(\arcB)$ for every cut arc $\arcB$ from the left figure.
The range in which we search in order to find the arc $\arcA$ intersecting $\cut'$ is shown in gray.}
\label{fig:improper}
\end{figure}

In this section, we show how the algorithm can be implemented so
that it uses $O(n\log n)$
time in total and $O(n)$ space when the input $\PP_1$ is a simply-connected curvilinear region.
The main difference between the general and the simply-connected case is that while a cut in a simply-connected region might split the region into many connected components (as seen in Figure~\ref{fig:improper}), these will have only one new cut arc each.
In particular, all new cut arcs that are not perfect will be the active cut arcs in their components.

The algorithm for general curvilinear regions described in Section~\ref{anImplementation:general} uses $O(n^2)$ time just to find the problematic objects to be handled.
This can be reduced to $O(n)$ time by storing pointers to the problematic objects on a stack $\stack$.
The algorithm starts by traversing the boundary of the original input $\PP_1$ and adds pointer to all problematic objects to $\stack$.
During the executing of the algorithm, we maintain $\stack$ so that it contains pointers to all problematic objects of $\PP$.
In particular, when an active cut arc is made, a pointer to the arc is pushed to $\stack$.
This ensures that the active cut arc will be handled first in its component.

We implement the stack $\sigma$ as a doubly linked list, so that elements in the middle of $\stack$ can be removed.
The cut performed due to one problematic object $\se_1$ popped from $\stack$ may result in another problematic object $\se_2\in\stack$ to disappear from $\partial\PP$.
By careful use of pointers from objects of $\PP$ to elements in $\stack$, we can remove pointers in $\stack$ that point to objects of $\PP$ that we remove in such situations.
In this way, we can ensure that when we pop a pointer to an object $\sigma$, the object $\sigma$ has not disappeared from $\partial\PP$ after the pointer to $\sigma$ was pushed to $\stack$.
Likewise, we update pointers to arcs that shrink due to a cut.
It may also happen that there is a pointer in $\stack$ to an arc $\arcA$ of $\PP$ that is split in two smaller arcs $\arcA_1$ and $\arcA_2$ due to a cut.
We may then update $\stack$ accordingly by replacing the old pointer to $\arcA$ with pointers to $\arcA_1$ and $\arcA_2$.

In order to get down to the promised running time of $O(n\log n)$, it then remains to improve the running time used on the subroutine $\ttt{TraverseC}$ from $O(n^2)$ to $O(n\log n)$.
Cheng, Cheong, Everett, and van Oostrum~\cite{Cheng04} described an efficient solution to the following problem.
For a simple polygon $V$, preprocess $V$ such that queries of the following
kind can be answered efficiently: Given a circular unit radius arc $\arcA$ beginning
at some point in the interior of $V$, find the first intersection point between $\arcA$ and
the boundary $\partial V$ if it exists.
The algorithm requires $O(n)$ space, uses $O(n\log n)$ time on preprocessing, and answers queries in $O(\log n)$ time, where $n$ is the number of vertices of $V$.

It is straightforward to generalize the method to work for curvilinear polygons.
We apply the preprocessing to the original input $\PP_1$.
Thus, by querying an arc following $\cut$ from $a$ in the direction through $\PP$, we know the point where $\cut$ exits $\PP_1$.
However, the arc $\cut$ can exit $\PP$ before it exits $\PP_1$, namely if and only if it crosses a cut arc of $\PP$.
In the following, we show how to detect if that is the case or not.

Using the circular ray shooting data structure in both directions from $a$ along $\cut$, we can find a maximal arc $\cut'\subset\partial\cutdisk$ such that $a\in\cut'$, $\cut'\subset\PP_1$, and $\cut'\cap \partial\PP_1=\{\start(\cut'),\terminal(\cut')\}$.
Note that if $a\in\partial\PP_1$, then $\start(\cut')=a$, but it is possible that $a$ is a point on a cut arc introduced by the algorithm so that $\start(\cut')\neq a$.
The following lemma says that if the arc $\cut'$, when traversed forward from $a$, enters a removed area, i.e., a connected component of $\PP_1\setminus\PP$, then it stays in that removed area.

\begin{lemma}\label{charArcB}
If $\cut'(a,\terminal(\cut')]$ leaves $\PP$ through a point $x$ on a cut arc $\arcA$, then $\cut'$ does not again enter $\PP$ after $x$, i.e., $\cut'(x,\terminal(\cut')]\subset\PP_1\setminus\PP$.
Similarly, we have $\cut'[\start(\cut'),a)\subset\PP_1\setminus\PP$.
\end{lemma}

\begin{proof}
We consider the first iteration $i$ such that $\cut'$ leaves $\PP_{i+1}$ and enters $\PP_{i+1}$ again later, when traversed from $\start(\cut')$.
If there are no such iterations, we conclude that $\cut'(x,\terminal(\cut')]\subset\PP_1\setminus\PP$, as stated.
Let $\cut_i$ be the arc that defined the cut made in $\PP_i$ in order to get $\PP_{i+1}$.
It follows that $\cut'$ crosses $\cut_i$ at a point $y$ and later again at a point $z\in \cut_i(y,\terminal(\cut_i)]$.
Note that the circle containing $\cut'$ is a curve $\gamma$ of bounded convex curvature.
Furthermore, it holds for the disk $\cutdisk_i=\cutdisk(\cut_i)$ that $\cutdisk_i$ is contained in the region bounded by $\gamma[y,z]\cup \partial\cutdisk_i(z,y)$.
We get from Lemma~\ref{lemma:algCurve} that $\gamma[y,z]\not\subset \cutreg(\cut_i)$.
But them $\cut'$ leaves and reenters $\PP_i$, which contradicts the minimality of $i$.

That $\cut'[\start(\cut'),a)\subset\PP_1\setminus\PP$ follows from an analogous argument.
\end{proof}

We say that a cut arc $\arcA$ is \emph{proper} if $\start(\arcA),\terminal(\arcA)\in\partial\PP_1$.
Otherwise, $\arcA$ is \emph{improper}; see Figure~\ref{fig:improper} (left) for an example where improper cut arcs are born, one of them being the arc $\cut'$.

\begin{lemma}\label{lem:improper}
In each connected component $\KO$ of $\PP$, if there are any improper cut arcs of $\KO$, then it is the active cut arc of $\KO$ and one or both of the neighbours.
\end{lemma}

\begin{proof}
We prove that if the statement holds in the beginning of an iteration, then it also holds in the beginning of the next.
Consider a component $\KO$ and suppose first that no cut arc of $\KO$ is improper.
Suppose that we make a cut in $\KO$ defined by an arc $\cut$.
Recall that the new cut arcs created are some of the arcs $\cut_1,\ldots,\cut_t$ in $\cut\cap\PP$.
The only other arcs that change are the neighbours of these new cut arcs.
An improper arc is created when one of the new cut arcs $\cut_j$ has one or two endpoints on an existing cut arc, which is not an original arc of $\PP_1$.
Then $\cut_j$ will be pushed to $\stack$ and be the active cut arc in its component. 
It then holds that the only other improper cut arcs are one or both of the neighbours of $\cut_j$.

Suppose now that there is an improper cut arc of $\KO$ in the beginning of an iteration.
We assume inductively that the active cut arc $\se$ of $\KO$ is improper, as is one or both of its neighbours $\arcA$ and $\arcB$, but not other arcs of $\KO$.
Suppose that both $\arcA$ and $\arcB$ are improper. 
We then make a cut of type 3 and may assume that we cut along the arc $\cut$ from $\start(\arcA)$ to $\terminal(\se)$.
Let the new cut arcs thus created be $\cut_1,\ldots,\cut_t$.
Here, $\cut_t$ will have endpoint $\terminal(\cut_t)=\terminal(\se)$, so $\cut_t$ is improper and the active cut arc in its component, and $\arcB$ is an improper cut arc in the same component which is a neighbour to $\cut_t$, so the statement holds.
The argument for the components of the other arcs $\cut_1,\ldots,\cut_{t-1}$ is identical as the one we gave for the case where no cut arc of $\KO$ is improper.

The case that only one of $\arcA$ and $\arcB$ is improper is also similar.
\end{proof}

By Lemma~\ref{lem:improper}, we know that there are at most three improper cut arcs to take into account when implementing $\ttt{TraverseC}$.
It therefore remains to find intersections between $\cut'(a,\terminal(\cut')]$ and the proper cut arcs.

Suppose that $\cut'(a,\terminal(\cut')]$ intersects a proper cut arc $\arcA$.
Then $\cut'$ and $\arcA$ can have one or two intersection points.
We now argue that if they have two, the first must be $a$: There can be no intersection at $\cut'[\start(\cut'),a)$ since $\cut'[\start(\cut'),a)\in\PP_1\setminus\PP$ (by Lemma~\ref{charArcB}) and $\arcA\subset\partial\PP$.
Similarly, Lemma~\ref{charArcB} gives that if $\cut'(a,\terminal(\cut')]$ intersect at a point $x$, then $\cut'(x,\terminal(\cut')]\subset\PP_1\setminus\PP$, so there can be no other intersection point than $x$ on the part $\cut'(a,\terminal(\cut')]$.
Therefore, we can easily check if $\cut'$ intersects the same proper cut arc twice, as that arc must be the arc of $\PP$ containing the point $a$.
We therefore turn our attention to finding a proper cut arc intersecting $\cut'$ once, where the intersection is on the part $\cut'(a,\terminal(\cut')]$.
This leads to the following lemma.

\begin{lemma}\label{charArcB2}
Let $\arcA$ be a proper cut arc of $\PP$.
Assume for a point $a\in\cut'\cap\partial\PP$ that $a\notin\arcA$ and that $a$ is on the convex side of $\arcA$, that is, $a\in \partial\PP[\terminal(\arcA),\start(\arcA)]$.
Then $\arcA$ and $\cut'(a,\terminal(\cut')]$ intersect if and only if the endpoints of $\cut'$ and $\arcA$ appear in the order $\start(\cut'),\start(\arcA),\terminal(\cut'),\terminal(\arcA)$ on $\partial\PP_1$.
\end{lemma}

\begin{proof}
If $\arcA$ and $\cut'$ intersect, then the intersection point is a unique point $x$ by Lemma~\ref{charArcB}, and $\cut'(x,\terminal(\cut')]\subset\PP_1\setminus\PP$.
Therefore, since $a\notin\arcA$, we know that $\arcA$ separates $\start(\cut')$ and $\terminal(\cut')$ in $\PP_1$.
As $a$ and $\start(\cut')$ are on the convex side of $\arcA$ and $\terminal(\cut')$ is on the concave side, this is equivalent to saying that the order is $\start(\cut'),\start(\arcA),\terminal(\cut'),\terminal(\arcA)$.
\end{proof}

Lemma~\ref{charArcB2} leads to our method for finding proper cut arcs intersecting
an arc $\cut'$ by searching after arcs with endpoints on specific portions of
$\partial\PP_1$.
We associate to each point $p$ on $\partial\PP_1$ a unique number
$\varphi(p)\in[0,n)$. Let the vertices of $\partial\PP_1$ be
$v_0,v_1,\ldots,v_{n-1},v_n$, where $v_0=v_n$. We set
$\varphi(v_i)\mydef i$ for $i<n$.
For the points $p$ on the arc between two vertices $v_i$ and
$v_{i+1}$, we interpolate between $i$ and $i+1$ to uniquely define $\varphi(p)$.
For a proper cut arc $\arcA$,
we define an associated point $\theta(\arcA)\in[0,n)\times(0,2n)$
in the following way:
$$
\theta(\arcA)\mydef
\begin{cases}
(\varphi(\start(\arcA)),\varphi(\terminal(\arcA))) & \textrm{if }\varphi(\start(\arcA))<\varphi(\terminal(\arcA)) \\
(\varphi(\start(\arcA)),\varphi(\terminal(\arcA))+n) & \textrm{otherwise}.
\end{cases}
$$

Lemma~\ref{charArcB2} then implies the following lemma, the use of which is demonstrated in Figure~\ref{fig:improper} (right).

\begin{lemma}\label{charArcB3}
Let $\arcA$ be a proper cut arc of $\PP$.
Assume that $a\in\cut'\cap\partial\PP$ and $a\notin\arcA$ and let $(x,y)\mydef \theta(\cut')$.
Then $\arcA$ and $\cut'(a,\terminal(\cut')]$ intersect
if and only if
\begin{itemize}
\item
$y<n$ and $\theta(\arcA)\in (x,y)\times (y,x+n)$, or

\item
$y\geq n$ and $\theta(\arcA)\in (x,n)\times (y,x+n)\cup [0,y-n)\times (y-n,x)$.
\end{itemize}
\end{lemma}

\begin{proof}
Lemma~\ref{charArcB2} says that $\arcA$ and $\cut'(a,\terminal(\cut')]$
intersect if and only if the endpoints of $\cut'$ and $\arcA$ appear in the order
$\start(\cut'),\start(\arcA),\terminal(\cut'),\terminal(\arcA)$
on $\partial\PP_1$.
If we assume that $y<n$ and $\varphi(\start(\arcA))<\varphi(\terminal(\arcA))$,
this is equivalent to
$\varphi(\start(\cut'))<\varphi(\start(\arcA))<\varphi(\terminal(\cut'))
<\varphi(\terminal(\arcA))<n$, which means that
$\theta(\arcA)\in (x,y)\times (y,n)$.
If $y<n$ and $\varphi(\start(\arcA))>\varphi(\terminal(\arcA))$, we get $\varphi(\start(\cut'))<\varphi(\start(\arcA))<\varphi(\terminal(\cut'))<n\leq
\varphi(\terminal(\arcA))<x+n$.
These two cases can then be expressed at once as $\theta(\arcA)\in (x,y)\times (y,x+n)$.

The case $y\geq n$ is handled in a similar way.
\end{proof}

For each proper cut arc $\arcA$ of $\PP$, we store
the point $\theta(\arcA)$ in a data structure $\Theta$.
It is necessary to add new points to and
delete points from $\Theta$ as the algorithm proceeds, since new
cut arcs are created and others become improper or completely removed.
We need to be able to find a point $\theta(\arcA)$ in a rectangle
specified by $\cut'$ as stated in Lemma~\ref{charArcB3}.
Therefore, we implement $\Theta$ as a fully dynamic orthogonal
range searching structure as described by Blelloch~\cite{blelloch2008space}.
Algorithm~\ref{traverseC} sketches how to implement $\ttt{TraverseC}$.

\begin{algorithm}[h]
\LinesNumbered
\DontPrintSemicolon
\SetArgSty{}
\SetKwInput{Input}{Input}
\SetKwInput{Output}{Output}
\SetKw{Report}{report}
\SetKwIF{If}{ElseIf}{Else}{if}{}{else if}{else}{end if}
\SetKwFor{Foreach}{for each}{}{end for}
\SetKwFor{While}{while}{}{end while}
Use the circular ray shooting data structure to find the arc
$\cut'$ such that $a\in\cut'$, $\cut'[\start(\cut'),\terminal(\cut')]\subset\PP_1$, and 
$\cut'\cap\partial\PP_1=\{\start(\cut'),\terminal(\cut')\}$.\;
Return the first intersection between $\cut(a,\terminal(\cut)]$ and $\partial\PP$ when following $\cut$ from $a$ by checking the following arcs:\;
\quad\quad The arc(s) containing $a$.\;
\quad\quad The improper cut arcs of $\partial\PP$, if any.\;
\quad\quad The arc represented by the point $\theta(\arcA)$ in the rectangle(s) as specified by Lemma~\ref{charArcB3}, if any.\;
\quad\quad The arc(s) containing $\terminal(\cut)$.
\caption{$\ttt{TraverseC}(\cut,a)$}
\label{traverseC}
\end{algorithm}

It is now possible to bound
the running time and memory requirement
of Algorithm~\ref{roundAlg} when using our suggested implementation.

\begin{theorem}\label{finalThm}
Algorithm~\ref{roundAlg} can be implemented so that it runs in
time $O(n\log n)$ and uses $O(n)$ space, assuming that the input is a simply-connected curvilinear region.
\end{theorem}

\begin{proof}
We first bound the space and time used on maintaining and querying the data structure
$\Theta$.
From Lemma~\ref{linearIts}, we know that there are $O(n)$ points in $\Theta$ at
any time. Blelloch~\cite{blelloch2008space} describes how to implement
$\Theta$ using $O(n)$ space so that insertions and deletions are performed in
$O(\log n)$ amortized time and
orthogonal range reporting queries in
$O\left(\log n+k\frac{\log n}{\log \log n}\right)$
time, where $k$ is the number of reported points.
In our case, due to Lemma~\ref{charArcB}, there are $0$ or $1$ points
to report in each query.
Therefore, we use $O(n\log n)$ time on $\Theta$ in total.

It takes $O(n\log n)$ time and uses $O(n)$ space to build the circular ray shooting data structure~\cite{Cheng04}.
For each cut arc of $\PP$, we perform $2$ circular ray shooting queries in order to find the arc $\cut'$ with endpoints on $\partial\PP_1$.
Thus, we use $O(n\log n)$ time building and querying the data structure in total.

The portion of $\partial\PP$ we traverse when calling
$\ttt{TraverseP}(\cut,b)$ is removed from $\PP$ and hence never traversed again.
Therefore, the time used on $\ttt{TraverseP}$ is bounded by the number of
vertices occuring on $\partial\PP$ during the execution of the algorithm.
By Lemma~\ref{linearIts}, we use $O(n)$ time on $\ttt{TraverseP}$
in total.
\end{proof}

\section{Concluding remarks}
It remains an interesting open problem if the running time of $O(n^2)$ for general curvilinear regions can be improved.
One way to do that would be to find an efficient data structure for circular ray shooting in polygonal domains and generalize the approach we used for simply-connected regions in Section~\ref{anImplementation:simply}.
In the case of straight line ray shooting in polygonal domains, Hershberger and Suri~\cite{hershberger95pedestrian} and Chazelle, Edelsbrunner, Grigni, Guibas, Hershberger, Sharir, and Snoeyink~\cite{chazelle94ray} devised data structures with preproccesing time $O(n\sqrt h+n\log n+h^{3/2}\log h)$ and query time $O(\sqrt h\log n)$, where $h$ is the number of holes (Chen and Wang~\cite{chen2015visibility} described a data structure with query time $O(\log n)$, at the cost of preprocessing time $O(n+h^2\,\text{polylog}\, n)$, which is quadratic in $n$ in the worst case).
This gives hope that a data structure for circular ray shooting in polygonal domains can be constructed where preprocessing and $O(n)$ queries together take $o(n^2)$ time.
But then we still need to find a way to replace the range searching data structure.
Another approach could be to find other cut rules that can be implemented more efficiently.

We finally mention a question of a purely mathematical nature:
For a set of points $S\subset\RR^2$, let $\mathcal B(S)$ be the family of subsets of $S$ of bounded convex curvature.
Is it true that for any set $S$, the union $\bigcup_{B\in\mathcal B(S)} B$ has bounded convex curvature?
In other words, is there a unique maximum subset of $S$ of bounded convex curvature?
It follows from our results that this holds when $S$ is a curvilinear region, but we expect it to be the case for all sets $S$.

\section{Data availability}

Data sharing is not applicable to this article as no datasets were generated or analyzed during the current study.

\section*{Acknowledgement}

We thank the anonymous reviewers for their thorough comments and one reviewer for spotting a glitch in the submitted manuscript.

\bibliography{bib}{}

\end{document}